
%
%
%
\input harvmac

\def\Titlehh#1#2{\nopagenumbers\abstractfont\hsize=\hstitle\rightline{#1}%
\vskip .2in\centerline{\titlefont #2}\abstractfont\vskip .2in\pageno=0}
\def\CTPa{\it Center for Theoretical Physics, Department of Physics,
      Texas A\&M University}
\def\CTPb{\it College Station, TX 77843-4242, USA}
\def\HARCa{\it Astroparticle Physics Group,
Houston Advanced Research Center (HARC)}
\def\HARCb{\it The Woodlands, TX 77381, USA}
\def\UAa{\it Department of Physics and Astronomy, University of Alabama}
\def\UAb{\it Box 870324, Tuscaloosa, AL 35487-0324, USA}
\def\CERN{\it CERN Theory Division, 1211 Geneva 23, Switzerland}
\def\ie{\hbox{\it i.e.}}     
\def\eg{\hbox{\it e.g.}}     

\def\nextline{\unskip\nobreak\hfill\break}
\def\coeff#1#2{{\textstyle{#1\over #2}}}

\catcode`\@=11 

\def\lsim{\mathrel{\mathpalette\@versim<}}
\def\gsim{\mathrel{\mathpalette\@versim>}}
\def\@versim#1#2{\vcenter{\offinterlineskip
    \ialign{$\m@th#1\hfil##\hfil$\crcr#2\crcr\sim\crcr } }}
\def\boxit#1{\vbox{\hrule\hbox{\vrule\kern3pt
      \vbox{\kern3pt#1\kern3pt}\kern3pt\vrule}\hrule}}

\def\etal{{\it et. al.}}
\def\r#1{$\bf#1$}
\def\rb#1{$\bf\overline{#1}$}

\def\t1{{\tilde 1}}
\def\ov{\overline}

\def\JL{J. L. Lopez}
\def\DVN{D. V. Nanopoulos}

\def\MeV{\,{\rm MeV}}
\def\GeV{\,{\rm GeV}}
\def\TeV{\,{\rm TeV}}

\def\wh{\widehat}
\def\wt{\widetilde}

\def\NPB#1#2#3{Nucl. Phys. B {\bf#1} (19#2) #3}
\def\PLB#1#2#3{Phys. Lett. B {\bf#1} (19#2) #3}

\def\PRD#1#2#3{Phys. Rev. D {\bf#1} (19#2) #3}
\def\PRL#1#2#3{Phys. Rev. Lett. {\bf#1} (19#2) #3}
\def\PRT#1#2#3{Phys. Rep. {\bf#1} (19#2) #3}
\def\MODA#1#2#3{Mod. Phys. Lett. A {\bf#1} (19#2) #3}

\def\TAMU#1{Texas A \& M University preprint CTP-TAMU-#1}

\nref\Carter{J. Carter, in Proceedings of the Joint International Lepton-Photon
Symposium and Europhysics Conference on High Energy Physics, Geneva,
Switzerland, 25 July--1 August 1991, ed. by S. Hegarty, K. Potter, and
E. Quercigh, p. 1 Vol. 2.}
\nref\Davier{M. Davier, as Ref. \Carter, p. 151 Vol. 2.}
\nref\topup{See \eg, J. Ellis, G. L. Fogli, and E. Lisi, \PLB{274}{92}{456};
F. Halzen, B. Kniehl, and M. L. Stong, University of Wisconsin preprint
MAD/PH/658 (Sep. 1991); F. del Aguila, M. Martinez, and M. Quiros, CERN
preprint CERN-TH.6389/92.}
\nref\tree{M. Veltmann, Acta Phys. Pol. B8 (1977) 475; C. Vayonakis,
Lett. Nuovo Cimento, {\bf17} (1976) 383;
B. W. Lee, C. Quigg, and H. B. Thacker, \PRD{16}{77}{1519};
M. L\"uscher and P. Weisz, \PLB{212}{89}{472}; W. Marciano, G. Valencia,
and S. Willenbrock, \PRD{40}{89}{1725}.}
\nref\loops{L. Durand, J. Johnson, and \JL, \PRL{64}{90}{1215} and
\PRD{45}{92}{3112}.}
\nref\LSZ{M. Lindner, M. Sher, and H. W. Zagluer, \PLB{228}{89}{139}.}
\nref\Sher{For a review see, M. Sher, \PRT{179}{89}{273}.}
\nref\EWx{K. Inoue, \etal, Prog. Theor. Phys. 68 (1982) 927; J. Ellis, \DVN,
and
K. Tamvakis, \PLB{121}{83}{123}; J. Ellis, J. Hagelin, \DVN, and K. Tamvakis,
\PLB{125}{83}{275}; L. Alvarez-Gaum\'e, J. Polchinski, and M. Wise,
\NPB{221}{83}{495}; L. Iba\~n\'ez and C. L\'opez, \PLB{126}{83}{54} and
\NPB{233}{84}{545}; C. Kounnas, A. Lahanas, \DVN, and M. Quir\'os,
\PLB{132}{83}{95}.}
\nref\KLNQ{C. Kounnas, A. Lahanas, \DVN, and M. Quir\'os, \NPB{236}{84}{438}.}
\nref\KLNPY{S. Kelley, \JL, \DVN, H. Pois, and K. Yuan, \PLB{273}{91}{423}.}
\nref\GRZ{G. Gamberini, G. Ridolfi, and F. Zwirner, \NPB{331}{90}{331}.}
\nref\minimal{For reviews see: R. Arnowitt and P. Nath, {\it Applied N=1
Supergravity} (World Scientific, Singapore 1983);
H. P. Nilles, \PRT{110}{84}{1}.}
\nref\noscale{E. Cremmer, S. Ferrara, C. Kounnas, and \DVN, \PLB{133}{83}{61};
J. Ellis, A. B. Lahanas, \DVN, and K. Tamvakis, \PLB{134}{84}{429};
J. Ellis, C. Kounnas, and \DVN, \NPB{241}{84}{406} and \NPB{247}{84}{373}.}
\nref\LN{For a review see A. B. Lahanas and D. V. Nanopoulos,
\PRT{145}{87}{1}.}
\nref\review{L. Ib\'a\~nez and G. G. Ross, CERN preprint CERN-TH.6412/91.}
\nref\CA{U. Amaldi, \etal, \PRD{36}{87}{1385}; G. Costa, \etal,
\NPB{297}{88}{244}.}
\nref\EKN{J. Ellis, S. Kelley and D. V.  Nanopoulos, \PLB{249}{90}{441},
\PLB{260}{91}{131}, \NPB{373}{92}{55}, and CERN preprint CERN-TH.6481/92.}
\nref\others{P. Langacker and M.-X. Luo, \PRD{44}{91}{817};
U. Amaldi, W. de Boer, and H. F\"urstenau, \PLB{260}{91}{447};
F. Anselmo, L. Ciafarelli, A. Peterman, and A. Zichichi, Nuovo Cim. {\bf104A}
(1991) 1817 and CERN preprint CERN-TH.6429/92;
R. Barbieri and L. Hall, \PRL{68}{92}{752}.}
\nref\Florida{H. Arason, \etal, \PRL{67}{91}{2933};
A. Giveon, L. Hall, and U. Sarid, \PLB{271}{91}{138}.}
\nref\RR{G. G. Ross and R. G. Roberts, Oxford preprint RAL-92-005.}
\nref\Shafi{B. Ananthanarayan, G. Lazarides, and Q. Shafi, \PRD{44}{91}{1613};
S. Kelley, \JL, and \DVN, \PLB{274}{92}{387}.}
\nref\DN{M. Drees and M.M. Nojiri, \NPB{369}{92}{54}.}
\nref\Japs{K. Inoue, M. Kawasaki, M. Yamaguchi, and T. Yanagida,
\PRD{45}{92}{328}.}
\nref\EZ{J. Ellis and F. Zwirner, \NPB{338}{90}{317}.}
\nref\Higgs{S. Kelley, \JL, \DVN, H. Pois, and K. Yuan, \TAMU{104/91} (to
appear
in Phys. Lett. B).}
\nref\ERZI{J. Ellis, G. Ridolfi, and F. Zwirner, \PLB{257}{91}{83}.}
\nref\DNh{M. Drees and M.M. Nojiri, \PRD{45}{92}{2482}.}
\nref\Drees{M. Drees and K. Hagiwara, \PRD{42}{90}{1709};
M. Drees, K. Hagiwara, and A. Yamada, \PRD{45}{92}{1725}.}
\nref\Nojiri{M.M. Nojiri, \PLB{261}{91}{76}.}
\nref\LNY{\JL, \DVN, and K. Yuan, \NPB{370}{92}{445} and \PLB{267}{91}{219}.}
\nref\ER{J. Ellis and L. Roszkowski, CERN preprint CERN-TH.6260/91.}
\nref\Japspd{M. Matsumoto, J. Arafune, H. Tanaka, and K. Shiraishi,
University of Tokyo preprint ICRR-267-92-5 (April 1992).}
\nref\AN{R. Arnowitt and P. Nath, \TAMU{24/92}.}
\nref\EENZ{J. Ellis, K. Enqvist, \DVN, and F. Zwirner, \MODA{1}{86}{57}.}
\nref\BG{R. Barbieri and G. Giudice, \NPB{306}{88}{63}.}
\nref\okada{Y. Okada, M. Yamaguchi, and T. Yanagida, Prog. Theor. Phys.
{\bf85} (1991) 1 and \PLB{262}{91}{54}.}
\nref\HH{H. Haber and R. Hempfling, \PRL{66}{91}{1815}.}
\nref\Falk{See \eg, N. K. Falck, Z. Phys. C {\bf30} (1986) 247, and references
therein.}
\nref\Polish{P. Chankowski, S. Pokorski, and J. Rosiek, \PLB{274}{92}{191};
A. Yamada, \PLB{263}{91}{233}.}
\nref\Brignole{A. Brignole, \PLB{277}{92}{313} and \PLB{281}{92}{284}.}
\nref\CW{S. Coleman and E. Weinberg, \PRD{7}{73}{1888}.}
\nref\W{S. Weinberg, \PRD{7}{73}{2887}.}
\nref\LS{S. Y. Lee and A. M. Sciaccaluga, \NPB{96}{75}{435}.}
\nref\Miller{R. D. C. Miller, \PLB{124}{83}{59} and \NPB{228}{83}{316}.}
\nref\DR{W. Siegel, \PLB{84}{79}{193}; D. M. Capper, D. R. T. Jones, and
P. Van Nieuwenhuizen, \NPB{167}{80}{479}.}
\nref\Jackiw{R. Jackiw, \PRD{9}{74}{1686}.}
\nref\Kang{J. S. Kang, \PRD{10}{74}{3455}.}
\nref\IP{J. Iliopoulos and N. Papanicolaou, \NPB{105}{76}{77}.}
\nref\K{B. Kastening, UCLA preprint UCLA/91/TEP/53 (December 1991).}
\nref\Russians{See \eg, K. G. Chetyrkin, \etal, \PLB{132}{83}{351}.}
\nref\GH{J. Gunion and H. Haber, \NPB{272}{86}{1}.}
\nref\SISM{S. Kelley, \JL, and \DVN, \PLB{278}{92}{140}.}
\nref\CEN{B. Campbell, J. Ellis, and \DVN, \PLB{141}{84}{229}.}
\nref\dt{H. P. Nilles, M. Srednicki, and D. Wyler, \PLB{124}{83}{337};
A. B. Lahanas, \PLB{124}{83}{341}; J. Kubo and S. Sasakibara,
\PLB{140}{84}{223}.}
\nref\MNTY{A. Masiero, \DVN, K. Tamvakis, and T. Yanagida, \PLB{115}{82}{380}.}
\nref\Barr{S. Barr, \PLB{112}{82}{219}, \PRD{40}{89}{2457}; J. Derendinger,
J. Kim, and \DVN, \PLB{139}{84}{170}.}
\nref\AEHN{I. Antoniadis, J. Ellis, J. Hagelin, and \DVN, \PLB{194}{87}{231}.}
\nref\aspects{J. Ellis, J. Hagelin, S. Kelley, and \DVN, \NPB{311}{88/89}{1}.}
\nref\decisive{I. Antoniadis, J. Ellis, J. Hagelin, and \DVN,
\PLB{231}{89}{65}; J. L. Lopez and \DVN, \PLB{251}{90}{73} and
\PLB{268}{91}{359}.}
\nref\SMCY{B. Greene, K.H. Kirklin, P.J. Miron, and G.G. Ross,
\PLB{180}{86}{69}; \NPB{278}{86}{667}; \NPB{292}{87}{606}; S. Kalara and
R.N. Mohapatra, \PRD{36}{87}{3474}; J. Ellis, K. Enqvist, \DVN, and K. Olive,
\NPB{297}{88}{103}; R. Arnowitt and P. Nath, \PRD{39}{89}{2006}.}
\nref\SMOrb{L. Ib\'a\~nez, H. Nilles, and F. Quevedo, \PLB{187}{87}{25};
L. Ib\'a\~nez, J. Mas, H. Nilles, and F. Quevedo, \NPB{301}{88}{157};
A. Font, L. Ib\'a\~nez, F. Quevedo, and A. Sierra, \NPB{331}{90}{421}.}
\nref\SMFFF{A. Faraggi, \DVN, and K. Yuan, \NPB{335}{90}{347};
A. Faraggi, \PLB{278}{92}{131};
K. Yuan, PhD Thesis Texas A\&M University (1991).}
\nref\Kap{V. Kaplunovsky, \NPB{307}{88}{145}; J. Derendinger, S. Ferrara,
C. Kounnas, and F. Zwirner, \NPB{372}{92}{145}.}
\nref\EN{J. Ellis and \DVN, \PLB{110}{82}{44}.}
\nref\Lacaze{I. Antoniadis, J. Ellis, R. Lacaze, and \DVN, \PLB{268}{91}{188};
S. Kalara, \JL, and \DVN, \PLB{269}{91}{84}.}
\nref\price{I. Antoniadis, J. Ellis, S. Kelley, and \DVN, \PLB{272}{91}{31}.}
\nref\DL{L. Durand and \JL, \PLB{217}{89}{463} and \PRD{40}{89}{207}.}
\nref\EHNOS{J. Ellis, J. S. Hagelin, D. V. Nanopoulos, K. A. Olive
and M. Srednicki, \NPB{238}{84}{453}.}
\nref\ENRS{K. Griest and J. Silk, Nature {\bf343} (1990) 26; L. Krauss,
\PRL{64}{90}{999}; J. Ellis, \DVN, L. Roszkowski, and D. Schramm,
\PLB{245}{90}{251}.}
\nref\chargino{L3 Collaboration, B. Adeva \etal, \PLB{233}{89}{530};
ALEPH Collaboration, D. Decamp \etal, \PLB{236}{90}{86};
OPAL Collaboration, M. Z. Akrawy \etal, \PLB{240}{90}{261}.}
\nref\CDF{CDF Collaboration, F. Abe \etal, \PRL{68}{92}{447}.}
\nref\PDG{Particle Data Group, \PLB{239}{90}{1}.}
\nref\top{CDF Collaboration, F. Abe, \etal, \PRD{43}{91}{664}.}
\nref\widths{V. Ruhlmann talk given at the LP-HEP Conference, Geneva, July
1991.}
\nref\ALHiggs{ALEPH Collaboration, D. Decamp \etal, \PLB{265}{91}{475};
P. Igo-Kemenes (OPAL Collaboration), as Ref. \Carter, p. 342 Vol.1;
L. Barone and S. Shevchenko (L3 Collaboration), as Ref. \Carter, p. 348 Vol.1.}
\nref\spart{A. Faraggi, J. Hagelin, S. Kelley, and \DVN, \PRD{45}{92}{3272}.}
\nref\KZ{Z. Kunszt and F. Zwirner, CERN-TH.6150/91, ETH-TH/91-7.}
\nref\Barger{V. Barger, M. S. Berger, A. L. Stange, and R. Phillips,
\PRD{45}{92}{4128}; H. Baer, M. Bisset, C. Kao, and X. Tata, FSU-HEP-911104,
UH-511-732-91, Nov. 1991; R. Bork, J. F. Gunion, H. E. Haber, and A. Seiden,
UCD-91-29, SCIPP-91/34, Nov. 1991; J. F. Gunion and L. H. Orr, UCD-91-15, Nov.
1991.}
\nref\BERZ{A. Brignole, J. Ellis, G. Ridolfi, and F. Zwirner,
\PLB{271}{91}{123}.}

\nfig\fI{Evolution with the renormalization scale $Q/M_Z$ of the
Higgs potential for a particular point in parameter
space (with $m_t=100\GeV$, $\tan\beta=3$, $m_{1/2}=150\GeV$, $m_0=A=0$, and
$\mu>0$). The various approximations used are: $V_0$ tree-level,
$\wt V_1$ one-loop subtracted, and $V_1$ one-loop unsubtracted.
Only $\wt V_1$ is $t$-independent to one-loop.}
\nfig\fII{Detail of the evolution of the minimum of the potentials shown in
\fI\ in the linear approximation in $t=\ln(Q/M_Z)$. Note the null slope of the
full one-loop subtracted potential ($\wt V_1$) and the non-zero
slope when only the dominant contributions to $\Delta V$ (\ie, $t,\tilde t,
b,\tilde b$) are included. Note also the non-zero slope of $\wt V'_1$ when
the masses in $\Delta V$ are allowed to run with $t$.}
\nfig\fIII{The scaled fine-tuning coefficient $\hat c_\mu$ as a function of
$\xi_0$ for various values of $m_t$ (and $\tan\beta=5$, $\mu>0$) for
(a) $\xi_A=0$, and (b) $\xi_A=\xi_0$. The middle curve corresponds to
$m_t=160\GeV$. Values of
$c_\mu=\hat c_\mu(m_{1/2}/M_Z)^2$ above $\Delta=100$ correspond to fine-tuning
of parameters to more than two orders of magnitude.}
\nfig\fIV{Same as \fIII\ but for the scaled fine-tuning coefficient
$\hat c_t$.}
\nfig\fV{Tree-level (solid) and one-loop (dashed) boundaries with {\it no}
phenomenological cuts imposed for (a) $\xi_0=0$, (b) $\xi_0=1$, and $\xi_A=0$,
$m_{1/2}=150,250\GeV$, and both sign of $\mu$ for the SSM. The dotted lines
are one-loop contours of $\mu$ ($\mu>0$ boundaries) and $B$ ($\mu<0$
boundaries) in GeV.}
\nfig\fVI{One-loop boundaries for the SSM with all consistency and
phenomenological cuts imposed for $m_{1/2}=150,250\GeV$, both
signs of $\mu$ and the following $(\xi_0,\xi_A)$ values: (a) $(0,-1)$ (dashed),
$(0,0)$ (solid), $(0,1)$ (dotted); (b) $(1,-1)$ (dashed), $(1,0)$ (solid),
$(1,1)$ (dotted). The figures show the progression of the left boundary to
higher values of $m_t$ due to the unit variations of $\xi_0$ and $\xi_A$.}
\nfig\fVII{One-loop boundaries for the SISM with all consistency and
phenomenological constraints imposed for $m_{1/2}=150,250\GeV$, both
signs of $\mu$ and two illustrative values of $(\xi_0,\xi_A)$: (a) $(0,0)$,
(b) $(1,0)$. The dotted lines are one-loop contours of $\mu$ ($\mu>0$
boundaries) and the lightest Higgs mass $m_h$ ($\mu<0$ boundaries) in GeV.}
\nfig\fVIII{The ratios of the first and second generation (a) slepton and
(b) squark masses relative to the gluino mass as a function of the gluino mass
in the SSM. The various lines represent the following: (a) $\tilde e_L$
(solid),
$\tilde e_R$ (dashed), $\tilde\nu$ (dotted);
(b) $\tilde u_L$ (solid), $\tilde u_R$ (dashed), $\tilde d_L$ (dotdash),
$\tilde d_R$ (dotted). When a splitting of lines of the same particle species
is noticeable (for the same $\xi_0$), this reflects the effect of the D-term
contributions to the sparticle masses (for $\tan\beta=2-10$).}
\nfig\fIX{One-loop boundaries for the SSM with all consistency and
phenomenological constraints imposed for $m_{1/2}=150,250\GeV$, both
signs of $\mu$ and two values of $(\xi_0,\xi_A)$: (a) $(0,0)$, (b) $(1,0)$.
The dotted lines are one-loop contours of $m_{\tilde\tau_1}$
($\mu>0$ boundaries) and the lightest Higgs mass $m_h$ ($\mu<0$ boundaries) in
GeV. Also shown are the corresponding $\tilde t_1$ mass ranges in GeV.}
\nfig\fX{One-loop boundaries for the SSM and SISM with all consistency and
phenomenological constraints imposed for $\mu>0$, $m_{1/2}=400\GeV$
(\ie, the fine-tuning upper bound), and $\xi_0=1,\xi_A=\pm1$. The dotted lines
are one-loop contours of $m_A$ in GeV. Within $5\%$ these can also be taken as
contours of $m_H$ and $m_{H^\pm}$.}

\Titlehh{\vbox{\baselineskip12pt\hbox{CERN-TH.6498/92}\hbox{CTP--TAMU--16/92}
\hbox{ACT--5/92}\hbox{UAHEP927}}}
{\vbox{\centerline{Aspects of Radiative Electroweak Breaking}
        \vskip2pt\centerline{in Supergravity Models}}}
\centerline{S. KELLEY$^{(a)(b)}$,
JORGE~L.~LOPEZ$^{(a)(b)}$, D.~V.~NANOPOULOS$^{(a)(b)(c)}$,}
\centerline{H. POIS$^{(a)(b)}$, and KAJIA YUAN$^{(d)}$}
\smallskip
\centerline{$^{(a)}$\CTPa}
\centerline{\CTPb}
\centerline{$^{(b)}$\HARCa}
\centerline{\HARCb}
\centerline{$^{(c)}$\CERN}
\centerline{$^{(d)}$\UAa}
\centerline{\UAb}
\vskip .1in
\centerline{ABSTRACT}
We discuss several aspects of state-of-the-art
calculations of radiative electroweak symmetry breaking in supergravity
models. These models have a five-dimensional parameter space in contrast
with the 21-dimensional one of the MSSM. We examine the Higgs one-loop
effective potential $V_1=V_0+\Delta V$, in particular how its
renormalization-scale ($Q$) independence is affected by the approximations
used to calculate $\Delta V$ and by the presence of a Higgs-field-independent
term which makes $V_1(0)\not=0$. We show that the latter must be subtracted out
to achieve $Q$-independence. We also discuss our own approach to the
exploration of the five-dimensional parameter space and the fine-tuning
constraints within this approach. We apply our methods to the determination of
the allowed region in parameter space of two models which we argue to be the
prototypes for conventional (SSM) and string (SISM) unified models. To this end
we impose the electroweak breaking constraint by minimizing the one-loop
effective potential and study the shifts in $\mu$ and $B$ relative to the
values
obtained using the tree-level potential. These shifts are most significant for
small values of $\mu$ and $B$, and induce corresponding shifts on the lightest
$\mu$- and/or $B$-dependent particle masses, \ie, those of the lightest stau,
neutralino, chargino, and Higgs boson states. Finally, we discuss the
predictions for the squark, slepton, and one-loop corrected Higgs boson masses.
\Date{May, 1992}

\newsec{Introduction}
The Standard Model of electroweak and strong interactions is well
established by now. In fact, the effects of the top quark in one-loop
electroweak processes predict its mass (within $\approx20\%$) centered
around $\approx140\GeV$ \Carter. Therefore, its expected direct experimental
detection in the near future will complete the set of Standard Model
predictions for the vector and fermion sectors. The scalar sector is another
story. The simplest electroweak symmetry breaking scenario with a single Higgs
boson is only mildly constrained experimentally,
with a lower bound of $m_H>57\GeV$ \Davier\ and no firm indirect experimental
upper bound, although this situation will change once the top quark mass is
measured \topup. On the other hand, interesting upper bounds on $m_H$ follow
from various theoretical assumptions, such as perturbative unitarity at
tree- ($m_H\lsim700\GeV$) \tree\ and one-loop ($m_H\lsim400\GeV$) \loops\
levels, and the stability of the Higgs potential ($m_H\lsim500\GeV$)
\refs{\LSZ,\Sher}.
In practice, with the advent of the SSC and LHC, experimental information about
the TeV scale is likely to clarify the composition of the Higgs sector.
Nevertheless, despite all these efforts the structure of the Standard Model and
its corresponding Higgs sector will remain basically unexplained.

It has therefore become
customary to turn to the physics at very high energies to search for answers
to these theoretical questions. The most promising theories of this kind
contain
two new ingredients: supersymmetry and unification. Together these can explain
the origin of the weak scale (\ie, the gauge hierarchy problem) relative to the
very high energy unification ($M_U$) or Planck ($M_{Pl}$) scales
\refs{\EWx,\KLNQ}. Furthermore, this class of theories predict a new set of
relatively light
($\lsim{\cal O}(1\TeV)$) particles consisting of partners for the Standard
Model particles but with spin offset by 1/2 unit. In fact, the new set of
particles appears ever more likely to overlap little with the mass scales of
the standard ones, thus their present unobserved status. Moreover, the
Standard Model Higgs boson will then appear as one of the new particles but
with mass close to $M_Z$, thus avoiding naturally the theoretical problems
mentioned above.

Unfortunately, the introduction of supersymmetry also increases significantly
the number of unknown parameters in the theory, mainly because this symmetry
must be softly broken at low energies. Indeed, to describe a generic
low-energy supersymmetric model (the so-called minimal supersymmetric
standard model (MSSM)) neglecting the first- and second-generation
Yukawa couplings, the KM angles, and possible CP violating phases,
we need the following set of parameters (the values
of $\sin^2\theta_w,\alpha_3,\alpha_e,M_Z$ are taken as measured parameters):
\item{{\bf a}.}The Yukawa ($\lambda_t,\lambda_b,\lambda_\tau$) and Higgs
mixing ($\mu$) superpotential couplings. (We can trade the Yukawa couplings
for $m_t,\tan\beta$; $m_b,m_\tau$, with $\tan\beta=v_2/v_1$ the ratio of
Higgs vacuum expectation values, and $m_b,m_\tau$ given.)
\item{{\bf b}.}The soft-supersymmetry breaking trilinear ($A_t,A_b,A_\tau$) and
bilinear ($B$) scalar couplings (corresponding to the superpotential couplings
in {\bf a}).
\item{{\bf c}.}The soft-supersymmetry breaking left-left and right-right
entries in the squark and slepton mass matrices for the first and second
($m_{Q,U^c,D^c},\,m_{L,E^c}$), and third
($m_{Q_3,U^c_3,D^c_3},\,m_{L_3,E^c_3}$)
generations.
\item{{\bf d}.}The soft-supersymmetry breaking gaugino masses $m_{\tilde g},
m_{\wt W},m_{\wt B}$.
\item{{\bf e}.}The Higgs sector parameter (at tree-level), \eg, the
pseudoscalar
Higgs boson mass $m_A$.

\noindent The above 21 unknown parameters make any thorough analysis of this
class of models rather impractical, and have allowed in the past only limited
explorations of this parameter space. If we now add the gauge unification
constraint ($\alpha_i(M_U)=\alpha_U,\,i=1,2,3$),
the assumption of universal soft-supersymmetry breaking at a
scale $\Lambda_{susy}=M_U$, and high-energy dynamics (in the form of
renormalization group equations (RGEs) for all the parameters involved), the
set of parameters in ${\bf b}$ reduces to $A=A_t=A_b=A_\tau$ and $B$, those
in $\bf c$ to $m_0=m_{Q,U^c,D^c}=m_{Q_3,U^c_3,D^c_3}=m_{L,E^c}=m_{L_3,E^c_3}$,
and those in $\bf d$ to $m_{1/2}=m_{\tilde g}=m_{\wt W}=m_{\wt B}$; these
relations are valid {\it only} at the scale $\Lambda_{susy}$. The number of
parameters has been dramatically reduced down to eight.

Let us now add low-energy dynamics by demanding radiative breaking of the
electroweak symmetry. The tree-level Higgs potential is given by
\eqna\Inti
$$\eqalignno{V_0=&(m^2_{H_1}+\mu^2)|H_1|^2+(m^2_{H_2}+\mu^2)|H_2|^2
                                                +B\mu(H_1H_2+{\rm h.c.})\cr
&+\coeff{1}{8}g^2_2(H^\dagger_2{\bf\sigma}H_2+H^\dagger_1{\bf\sigma}H_1)^2
+\coeff{1}{8}g'^2\left(|H_2|^2-|H_1|^2\right)^2,&\Inti{}\cr}$$
where $H_1\equiv{{H^0_1\choose H^-_1}}$ and $H_2\equiv{{H^+_2\choose H^0_2}}$
are the two complex Higgs doublet fields, $g'=\sqrt{5/3}\,g_1$ and $g_2$ are
the $U(1)_Y$ and
$SU(2)_L$ gauge couplings, and $B\mu$ is taken to be real and negative. This
potential has a minimum\foot{The parameters in $V_0$ must satisfy further
consistency constraints to insure that this is a true minimum of the tree-level
Higgs potential (see \eg, Ref. \KLNPY). As discussed in Ref. \GRZ\ (and below),
the one-loop effective potential satisfies most of these constraints
automatically.} if $\partial V_0/\partial\phi_i=0$, with $\phi_i$ denoting the
eight real degrees of freedom of $H_1$ and $H_2$. In particular, for
$\phi_i={\rm Re}\,H^0_i$ one obtains two constraints which allow the
determination of $\mu$ and $B$,
\eqna\Intii
$$\eqalignno{\mu^2&={m^2_{H_1}-m^2_{H_2}\tan^2\beta\over\tan^2\beta-1}
-\coeff{1}{2}M^2_Z,&\Intii a\cr
B\mu&=-\coeff{1}{2}\sin2\beta(m^2_{H_1}+m^2_{H_2}+2\mu^2)<0,&\Intii b\cr}$$
up to the sign of $\mu$. In these expressions, $m^2_{H_1},m^2_{H_2}$ are
soft-supersymmetry breaking masses equal to $m^2_0$ at $\Lambda_{susy}$.
Since the whole set of Higgs masses and couplings (at tree-level) follows
from $m^2_A$ (and $\tan\beta$), and one can easily show that
$m^2_A=-2B\mu/\sin2\beta$, the
parameter in $\bf e$ is also determined. (This result also holds at one-loop
although the expression for $m^2_A$ is more complicated in this case.)

The final parameter count in the class of models we consider is then just
five: $m_t,\tan\beta,m_{1/2},m_0,A$ (plus the sign of $\mu$). Note also that
$\sin^2\theta_w$ (as well as $M_U$ and $\alpha_U$) gets determined (from
$\alpha_3$ and $\alpha_e$) by the gauge unification condition. What are the
{\it a priori} expected values of $m_{1/2},m_0,A$? In principle choosing a
suitable supergravity model (\ie, suitable hidden sector) one could have
arbitrary values for these parameters. The so-called `minimal' supergravity
models \minimal\ predict $A=B+m_0$ (at $\Lambda_{susy}$) with
$m_0,m_{1/2}\not=0$ in general. Supergravity models which explain naturally
the vanishing of the cosmological constant (even after supersymmetry breaking),
the so-called {\bf no-scale} models \refs{\noscale,\LN}, typically require
$m_0=A=0$ and $m_{1/2}\not=0$, although models with $m_0,A\not=0$ are possible
also.

With this scenario in mind we have undertaken a systematic study of
supersymmetric unified models with universal soft-supersymmetry breaking.
We pursue this objective in three steps: (i) determination of the parameter
space allowed by all consistency and experimental constraints using the Higgs
one-loop effective potential;
(ii) calculation of particle masses, upper and lower bounds, correlations
among them, and discovery limits; and (iii) study of specific reactions
such as collider processes relevant for particle detection, rare
decays, neutralino dark matter, supersymmetric loop corrections to
$\sin^2\theta_w$, etc. In this paper we address the first two points.

Analyses of this nature in {\it unified} models exist in the literature
(although their number is small relative to those in {\it generic} low-energy
supersymmetric models) and date back one full decade (for a recent review
see \review). The recent interest in this type of analyses has been spurred
by the experimental data available from LEP, and it has mostly been concerned
with systematic studies of various aspects of unified supersymmetric models,
such as gauge coupling unification \refs{\CA,\EKN,\others,\Florida,\RR}, the
Yukawa sector \refs{\Florida,\Shafi}, electroweak symmetry breaking at
tree-level \refs{\DN,\Japs,\KLNPY,\RR} and one-loop \refs{\GRZ,\EZ,\Higgs}, the
one-loop corrected Higgs boson masses \refs{\ERZI,\DNh,\Higgs}, supersymmetry
loop effects on the $\rho-$parameter and $\sin^2\theta_w$ \refs{\Drees,\RR},
neutralino dark matter \refs{\EZ,\Nojiri,\LNY,\ER}, proton decay
\refs{\Japspd,\AN}, etc.

The purpose of this paper is to sharpen the determination of the allowed
parameter space of this class of models (obtained at tree-level in Ref.
\KLNPY) by using the one-loop Higgs effective potential. In Sec. 2
we compare both approximations and discuss several theoretical and practical
issues related to the use of the one-loop effective potential. We present
a numerical study of the effect of the usual approximations to the one-loop
effective potential (\eg, keeping only the top-stop contributions) on its
renormalization-scale ($Q$) independence properties. We also quantify the
phrase
``up to two-loop effects", clearly identifying the one- and two-loop leading-
and non-leading-log contributions, and obtain a $Q$-independent one-loop
potential. We show that in problems where the value of the one-loop effective
potential (as opposed to any of its field derivatives) is relevant,\foot{These
problems include studies of the cosmological constant and dynamical
determinations of supersymmetry breaking parameters in no-scale supergravity
models.} one must define a new `subtracted' potential which vanishes at the
origin of field space and is manifestly renormalization-scale independent.
We also provide details of our numerical and analytical techniques used to
calculate the one-loop values of $\mu$ and $B$ and to demonstrate one-loop
$Q$-independence.

In Sec. 3 we discuss the ``fine-tuning" constraint \refs{\EENZ,\BG}, which
limits the parameter space and yields an allowed region bounded in all five
variables. In Sec. 4 we show that the class of models we consider here
constitute a good first approximation to any realistic traditional (SSM) or
string (SISM) supersymmetric unified model below the unification scale.
In Sec. 5 we discuss the
consistency and phenomenological constraints on the parameter space and the
ensuing bounded regions in the $(m_t,\tan\beta)$ plane. We then study the
shifts in $\mu$ and $B$ obtained with the one-loop potential relative to their
tree-level counterparts, and find them to be most significant for small values
of $\mu$ and $B$. We also explore the allowed parameter space extending our
previous analysis to nonvanishing values of $m_0$ and $A$, and using the
one-loop effective potential. In Sec. 6 we study the squark and slepton masses
and the effect of the shifted values of $\mu$ and $B$ on the masses of the
lightest stau, the lightest neutralino, and the lightest chargino states. We
also study the one-loop corrected Higgs boson masses and present tables of mass
ranges for all particle species for typical values of the parameters. Finally,
in Sec. 7 we summarize our conclusions and in the Appendix we present some
details of the computation of the one-loop effective potential.
\newsec{The effective potential}
As a practical matter, in studies of electroweak symmetry breaking it is
important to determine whether the tree-level approximation to the scalar
Higgs potential ($V_0$) is adequate for the purposes at hand. There are
two observations which make it apparent that this approximation may not be
accurate enough nor reliable (at least in some regions of parameter space):
(i) It had
been noted long ago \KLNQ\ (and it has been emphasized recently \GRZ) that
the low-energy scale $Q$ at which the RGEs connecting physics at very high
energies with physics at the electroweak scale are stopped, influences the
(tree-level) physical predictions of this class of models in significant ways,
such as in the determination of minima of the tree-level Higgs potential.
Equivalently
put, $V_0$ does not obey the RGE $dV_0/dt=0$ (with $t=\ln Q$) in any sensible
approximation. (ii) It has been recently observed that the one-loop corrections
to the Higgs boson masses are non-negligible in certain regions of
parameter space (\eg, for large $m_t$) \refs{\okada,\ERZI,\HH}.

It has been argued \GRZ\ that the one-loop effective potential
$V_1=V_0+\Delta V$ solves the problem in (i) since it is presumed to satisfy
the corresponding RGE ($dV_1/dt=0$) automatically and therefore yields
one-loop $Q$-independent predictions. In fact, this property of $V_1$ is
routinely used to derive one-loop RGEs \Falk. However, the one-loop
$Q$-independence of $V_1$ has been explored only in a limited
way in Ref. \GRZ\ and the details by which it works have so far remained
unclear or at least not explicitly described. Additionally, it has been shown
that the naive expectation of using the one-loop Higgs potential to obtain
one-loop corrected Higgs masses is in fact accurate to
${\cal O}({\rm few}\GeV)$ \refs{\Polish,\Brignole}, and therefore one-loop
corrections to $V_0$ are expected to affect the values of $\mu$ and $B$
calculated (as described in the Introduction) from the corresponding
minimization conditions
$\partial V_1/\partial{\rm Re}\,H^0_1=\partial V_1/\partial{\rm Re}\,H^0_2=0$.
The expression for $\Delta V$ \Sher,
\eqn\I{\Delta V={1\over64\pi^2}{\rm STr}\,{\cal M}^4
\left(\ln{{\cal M}^2\over Q^2}-{3\over2}\right),}
where ${\rm STr}\,f({\cal M}^2)
=\sum_j(-1)^{2j}(2j+1){\rm Tr}\,f({\cal M}^2_j)$ and ${\cal M}^2_j$ are the
field--dependent spin-$j$ mass matrices, receives contributions from all
particle species. However, in analyses of this kind it is customary to include
only the ``dominant" contributions to $\Delta V$, \ie,  those from the top-stop
(and sometimes also bottom-sbottom) system(s).
This statement is indeed accurate provided one is only interested in the
field derivatives of $V_1$. However, it is not clear to what extent this
approximation to $\Delta V$ will affect the $Q$-independence of $V_1$. We will
show that this question is not only of conceptual but also of practical
importance. In the remainder of this section we review the derivation of
$V_1$, study its formal $Q$-independence properties (in a $\lambda\phi^4$
theory and the MSSM), and discuss the numerical methods used to obtain
the one-loop values of $\mu$ and $B$.
\subsec{General remarks}
The first calculation of the one-loop effective potential in field theories
was performed by Coleman and E. Weinberg \CW. Their method consisted basically
of summing the infinite series of one-loop diagrams with all possible numbers
of zero-momentum external lines. This procedure becomes very cumbersome for
two- and higher-loop calculations and is replaced in practice (even at
one-loop) by the so-called ``tadpole method" first noticed by S. Weinberg
\W\ and later developed by Lee and Sciaccaluga \LS\ and extended to
supersymmetric theories by Miller \Miller. This method is derived \Sher\ by
expanding the effective action around shifted values of the fields. The
first derivative of the $n$-loop contribution to the effective potential
($V^{(n)}$) is simply given by the $n$-loop tadpole ($\Gamma^{1}_{(n)}$)
diagrams calculated in the shifted theory and with zero external
momentum.\foot{In what follows we denote $V_n=V^{(0)}+V^{(1)}+\cdots+V^{(n)}$.
Note that $V_0=V^{(0)}$ and $\Delta V=V^{(1)}$.}
For example, consider a massless $\lambda\phi^4$ theory with scalar potential
$V_0(\phi)={\lambda\over4!}\phi^4$. Shifting $\phi\to\phi-w$ gives a linear
term in $\phi$, $-{\lambda\over6}\phi w^3$, a ``mass term"
${\lambda\over4}w^2\phi^2$, and cubic term $-{\lambda\over6}w\phi^3$. At
tree-level we have $\Gamma^{1}_{(0)}={\lambda\over6}w^3$ and therefore
$V^{(0)}(\phi)={\lambda\over4!}\phi^4$ as expected. The constant of integration
has been fixed such that $V^{(0)}(0)=0$. At one-loop we have a single diagram
due to the induced cubic coupling, giving
\eqn\II{\Gamma^{1}_{(1)}={1\over2}\int {d^4k\over(2\pi)^4}{\lambda w\over
k^2-{\lambda\over2}w^2},}
where the ${1\over2}$ is the symmetry factor for the one-loop diagram.
This integral is divergent and can be regulated by introducing a cut-off
or more conveniently by using dimensional regularization. The result using
the latter is
\eqn\III{{dV^{(1)}\over dw}=\Gamma^1_{(1)}={\lambda^2 w^3\over4(4\pi)^2}
\left[ {1\over\hat\epsilon}+\ln\left({\lambda w^2\over2Q^2}\right)-1\right],}
where $D=4-2\epsilon$, $1/\hat\epsilon=-1/\epsilon+\gamma_E-\ln4\pi$, and $Q$
is an arbitrary scale introduced for dimensional reasons. In the $\ov{MS}$
renormalization scheme we simply discard $1/\hat\epsilon$ and obtain the finite
result immediately by integrating Eq. \III,
\eqn\IV{V^{(1)}(Q,\phi)={m^4(\phi)\over64\pi^2}\left[
\ln\left({m^2(\phi)\over Q^2}\right)-{3\over2}\right],}
where we have defined a field-dependent mass
$m^2(\phi)\equiv{1\over2}\lambda\phi^2$ to make the connection with Eq. \I\
more apparent. The arbitrary
scale $Q$ appearing in $V^{(1)}$ can be specified in connection with physically
measured (or `renormalized') quantities. One usually sets $\partial^2 V_1/
\partial\phi^2=0$ to preserve the `masslessness' of the theory and
$\partial^4 V_1/\partial\phi^4 (\phi=M)=\lambda_R$. The latter relation
implies that the quartic coupling takes the value $\lambda_R$ at scale $Q=M$.
In this case one gets
\eqn\V{\lambda_R=\lambda+{24\lambda^2\over(16\pi)^2}
\left[\ln\left({\lambda M^2\over2Q^2}\right)+{8\over3}\right]}
and
\eqn\VI{V_1=V^{(0)}+V^{(1)}=
{\lambda_R\over4!}\phi^4+{\lambda^2_R\phi^4\over(16\pi)^2}
\left[\ln\left({\phi^2\over M^2}\right)-{25\over6}\right].}
Alternatively, one may just leave $Q$ unspecified as in Eq. \IV. This is
much more convenient in complicated theories (like the MSSM) but has the
drawback that it does not involve true physical parameters measured at a
specific scale. Until the sparticle spectrum is detected, high accuracy
predictions are not required (unlike \eg, the electroweak sector in the SM),
and this approximation is perfectly adequate.

The calculation of $V^{(1)}$ sketched above can be easily generalized to
gauge theories with fermions \Sher. The final result for the one-loop
contribution to the effective potential is as given in Eq. \I\ with
$V^{(1)}=\Delta V$. We should remark that this result has been obtained in
the convenient Landau gauge (and using the $\ov{DR}$ renormalization scheme
\DR, \ie, the supersymmetry-preserving counterpart of the usual $\ov{MS}$
scheme
wherein dimensional regularization is replaced by dimensional reduction) and
is otherwise gauge dependent \refs{\W,\Jackiw}. However, physical quantities
extracted from $V^{(1)}$ should be gauge independent. This was explicitly
verified to one-loop in one example in Ref. \Kang\ and shown generally to hold
to all orders in perturbation theory in Ref. \IP. Therefore the Landau gauge
choice is as good as any other one.
\subsec{$Q$-(in)dependence}
In perturbative expansions involving the renormalization group, it is well
known that cancellation of the $Q$-dependence takes place across different
orders. For example, the tree-level QCD cross section for jet production is
highly $Q$-dependent, whereas the one-loop corrected expression is
$Q$-independent up to two-loop effects. We now study the analogous effect
for the scalar Higgs potential.
It is instructive to study a particularly simple field theory in detail to see
in what sense and to what degree of approximation is $V_1=V^{(0)}+V^{(1)}$
$Q$-independent. A calculation analogous to the one sketched above gives for
the massive $\lambda\phi^4$ theory
\eqn\VII{V_1(Q,\phi)={1\over2}\mu^2\phi^2+{\lambda\over4!}\phi^4+
{m^4(\phi)\over64\pi^2}
\left[\ln\left({m^2(\phi)\over Q^2}\right)-{3\over2}\right],}
with $m^2(\phi)\equiv\mu^2+{1\over2}\lambda\phi^2$.
To study the $Q$-dependence of $V_1$ we take the total derivative
$dV_1/dt$ and make use of the one-loop RGEs $d\lambda/dt=3\lambda^2/(4\pi)^2$,
$d\mu^2/dt=\mu^2\lambda/(4\pi)^2$, and $d\phi/dt=0$, to obtain
\eqn\VIII{{dV_1\over dt}=-{\mu^4\over2(4\pi)^2}+``{\rm two-loop}",}
where $\rm ``two-loop"$ denotes contributions where the derivative has acted
on pieces in $V^{(1)}$ other than $\ln Q^2$. Using the RGEs one can easily
see that these pieces are of higher order (\eg, $\propto1/(4\pi)^4$ as
opposed to $\propto1/(4\pi)^2$) and will be cancelled by two-loop contributions
in $V^{(2)}$. Equation \VIII\ indicates that $V_1$ does {\it not} obey the
RGE $dV_1/dt=0$ in one-loop approximation. This problem can be easily traced
back to the constant piece $V_1(Q,0)=V^{(1)}(Q,0)$ which may be subtracted from
$V_1$ such that
\eqna\IX
$$\eqalignno{\wt V_1(Q,\phi)&\equiv V_1(Q,\phi)-V_1(Q,0)&\IX a\cr
&=V^{(0)}(\phi)+V^{(1)}(Q,\phi)-V^{(1)}(Q,0),&\IX b\cr}$$
satisfies $\widetilde V_1(Q,0)=0$. Indeed, in this case
$V_1(Q,0)=\mu^4/(64\pi^2)[\ln(\mu^2/Q^2)-3/2]$ and
\eqn\X{{dV_1(Q,0)\over dt}=-{\mu^4\over2(4\pi)^2}+``{\rm two-loop}",}
such that $d\wt V_1/dt=0+``{\rm two-loop}"$ obeys the one-loop RGE as it
should. The above prescription (Eq. \IX) can be justified in several ways
(see also the recent discussion in Ref. \K). First of all, from the practical
point of view, adding a field-independent piece to $V$ is perfectly harmless
in problems where only field derivatives of $V$ are of interest. From the
requirement of renormalizability we know that the all-orders solution to the
RGE $dV/dt=0$ must
resemble $V^{(0)}$ but with suitably modified coefficients and wave-function
renormalizations, and this expression clearly vanishes at the origin of field
space. Yet another way of seeing this is in the $\ov{MS}$ cancellation of
infinities process. The counterterms $\propto\phi^2/\hat\epsilon$ and
$\propto\phi^4/\hat\epsilon$ that need to be added to the Lagrangian will not
absorb the $\propto\mu^4/\hat\epsilon$ divergence (see \eg, Eq. \VII\ with
$1/\hat\epsilon$ added inside the square brackets) and one is forced to
add a `constant' counterterm $\propto1/\hat\epsilon$ to remove it. This
occurs naturally for $\wt V_1$. In the context of supersymmetric theories, the
fact that $V(0)=0$ to all orders is equivalent \Miller\ to the statement that
unbroken (global) supersymmetry at tree-level cannot be broken by radiative
corrections.

Does this `subtraction' procedure survive higher-loop corrections? To study
this question we consider the two-loop effective potential $V_2=V^{(0)}
+V^{(1)}+V^{(2)}$, with $V^{(2)}$ given by \K
\eqna\XI
$$\eqalignno{V^{(2)}={1\over(4\pi)^4}\Biggl\{
&4\lambda m^4\left[\coeff{1}{32}\ln^2{m^2\over Q^2}
-\coeff{1}{16}\ln{m^2\over Q^2}+a_{200}\right]\cr
&+2\lambda^2\phi^2m^2\left[\coeff{1}{16}\ln^2{m^2\over Q^2}
-\coeff{1}{4}\ln{m^2\over Q^2}+a_{210}\right]\Biggr\},&\XI{}\cr}$$
where $m^2\equiv \mu^2+{\lambda\over2}\phi^2$ and $a_{200},a_{210}$ are some
numerical coefficients. We also use the following two-loop expressions for the
relevant RGEs \Russians
\eqna\XII
$$\eqalignno{{d\lambda\over dt}&={3\lambda^2\over(4\pi)^2}
                                -{(17/3)\lambda^3\over(4\pi)^4},&\XII a\cr
{1\over\mu^2}{d\mu^2\over dt}&={\lambda\over(4\pi)^2}
                                -{(5/6)\lambda^2\over(4\pi)^4},&\XII b\cr
{1\over\phi}{d\phi\over dt}&=-{(1/12)\lambda^2\over(4\pi)^4}.&\XII c\cr}$$
After some straightforward algebraic manipulations and keeping only terms
up to two-loop order we obtain
\eqna\derivs
$$\eqalignno{
{dV^{(0)}\over dt}&={m^4\over2(4\pi)^2}-{\mu^4\over2(4\pi)^2}
                        -{\lambda^2\phi^2 m^2\over2(4\pi)^4},&\derivs a\cr
{dV^{(1)}\over dt}&=-{m^4\over2(4\pi)^2}
                        +{\lambda m^2\over2(4\pi)^4}(m^2+\lambda\phi^2)
                                \left(\ln{m^2\over Q^2}-1\right),&\derivs b\cr
{dV^{(2)}\over dt}&=-{\lambda m^2\over2(4\pi)^4}(m^2+\lambda\phi^2)
                \left(\ln{m^2\over Q^2}-1\right)
                        +{\lambda^2\phi^2 m^2\over2(4\pi)^4},&\derivs c\cr}$$
and therefore
\eqn\XIII{{d\over dt}(V^{(0)}+V^{(1)}+V^{(2)})=-{\mu^4\over2(4\pi)^2}
                                        +``{\rm three-loop}".}
That is, the one-loop subtraction in Eq. \IX{} is still needed at two-loops
as anticipated. However, an analogous two-loop subtraction is not necessary
since the field-independent pieces in $dV^{(1)}/dt$ and $dV^{(2)}/dt$
($\propto\lambda\mu^4(\ln\mu^2/Q^2-1)/(4\pi)^4$) cancel among themselves. This
is a rather interesting result which may be not that surprising if one realizes
that in the cancellations of the field-independent pieces in $dV^{(n)}/dt$
and $dV^{(n+1)}/dt$, there is a mismatch for $n=0$ since $dV^{(0)}/dt$ has no
field-independent piece. It is not clear to us whether this phenomenon
persists at higher orders or in more complicated theories.

We now consider the case of the MSSM (our unification constraints do not affect
the present discussion). Analogous to the massive $\lambda\phi^4$ theory, we
define a potential $\wt V_1=V_0+\Delta V-\Delta V(0)$ for the MSSM using
Eqs. \Inti{} and \I. It is then not hard to see that the $Q$-dependence of
$\wt V_1$ can be studied to one-loop order from the following expression
\eqn\XIV{{d\wt V_1\over dt}={dV_0\over dt}-{1\over32\pi^2}{\rm STr}
\left\{{\cal M}^4(h_1,h_2)-{\cal M}^4(0,0)\right\}+``{\rm two-loop}",}
where $h_{1,2}\equiv {\rm Re}\,H^0_{1,2}$. This equation evidences the fact
that it is
the ${\rm STr}{\cal M}^4$ term in $\Delta V$ which cancels the running with
$Q$ of the parameters in $V_0$. It is also clear that an incomplete set of
contributions to the supertrace will make for an `incomplete' $Q$-independence
of $\wt V_1$. Below we quantify this statement, but first we consider a
particularly simple limit of the MSSM where the derivative in Eq. \XIV\ can
be carried out explicitly, and justifies our use of the subtracted one-loop
potential.

Let us take the limit of vanishing gauge couplings and $\mu=A_{t,b,\tau}=
\lambda_{b,\tau}=0$ (see Ref. \ERZI). This allows us to set $v_1=0$
and the tree-level potential reduces to $V_0=m^2_{H_2}h^2_2$.  The
relevant RGEs for the left-over tree-level parameters are
\eqna\XV
$$\eqalignno{
{dm^2_{H_2}\over dt}&={3\over8\pi^2}\lambda^2_t(2m^2_{\tilde q}+m^2_{H_2}),
                                                                &\XV a\cr
{dh^2_2\over dt}&=-{3h^2_2\over8\pi^2}\lambda^2_t,&\XV b\cr}$$
once we take all squark mass parameters degenerate
($m_Q=m_{U^c}=\cdots=m_{\tilde q}$). Therefore
\eqn\XVI{{dV_0\over dt}={3m^2_t\over8\pi^2}(2m^2_{\tilde q}+m^2_{H_2})
-{3m^2_tm^2_{H_2}\over8\pi^2}={3m^2_{\tilde q}m^2_t\over4\pi^2}.}
(Note the role played in the result by the often-neglected `vev running'.)
With our approximations only the top-stop contribution
($m^2_t=\lambda^2_t h^2_2,\ m^2_{\tilde t}=m^2_{\tilde q}+\lambda^2_t h^2_2$)
to Eq. \XIV\ survives, since any field-independent contribution is subtracted
out and with vanishing gauge couplings this is the case for the gauge/gaugino
and Higgs/Higgsino sectors. We then get
\eqn\XVII{{\rm STr}\left\{{\cal M}^4(h_1,h_2)-{\cal M}^4(0,0)\right\}\to
12\left\{\left[(m^2_t+m^2_{\tilde q})^2-m^4_t\right]-m^4_{\tilde q}\right\}
=24m^2_tm^2_{\tilde q}.}
Substituting Eqs. \XVI\ and \XVII\ into \XIV\ finally gives $d\wt V_1/dt=0$
as expected. Note that had we not used the subtracted one-loop potential
we would have obtained $d\wt V_1/dt=-3m^4_{\tilde q}/8\pi^2$, \ie, a huge
$Q$-dependence (see below).

Let us now take a more quantitative look at the $Q$-dependence of the various
approximations to the effective potential. To this end we expand $V_0$ and
$\wt V_1$ around $t=0$ ($t\equiv\ln(Q/M_Z)$), as follows
\eqn\XVIII{V_0(t,\phi)=a_0+a_1t+a_2t^2+a_3t^3\cdots,\qquad
                                        a_n\equiv{d^nV_0\over dt^n}(t=0,\phi),}
and
\eqn\XIX{\wt V_1(t,\phi)=V_0(t,\phi)+\Delta V(t,\phi)-\Delta V(t,0)
                                =b_0+b_1t+b_2t^2+b_3t^3+\cdots,}
with
\eqn\XX{\Delta V(t,\phi)=\coeff{1}{64\pi^2}{\rm Str}\,{\cal M}^4
\left(\ln{{\cal M}^2\over Q^2}-{3\over2}\right)(t,\phi).}
To one-loop order we only need to include in $\Delta V(t,\phi)$ the tree-level
masses evaluated at $t=0$ ($Q=M_Z$).
In \fI\ we plot the the tree-level ($V_0$) and one-loop ($\wt V_1$)
Higgs potentials (following a numerical procedure described in Sec. 2.3)
as a function of $t=\ln(Q/M_Z)$ for a particular point in parameter space (with
$m_t=100\GeV$, $\tan\beta=3$, $m_{1/2}=150\GeV$, $m_0=A=0$, $\mu>0$). (Note the
large $t$-dependence of the unsubtracted one-loop effective potential $V_1$,
as anticipated.) In calculating ${\rm Str}{\cal M}^{2,4}$
we have included the {\it full} spectrum. The curves shown are well fit by
third-degree polynomials in $t$, as follows\foot{The values for the various
potentials have been rescaled by $(m_{1/2})^4$.}
\eqna\XXI
$$\eqalignno{
a_0&=-0.0795,\quad a_1=0.2751,\quad a_2=-0.0315,\quad a_3=0.0027,&\XXI a\cr
b_0&=-0.0895,\quad b_1=-0.0007,\quad b_2=-0.0322,\quad b_3=0.0027.&\XXI b\cr}$$
(The non-vanishing $a_{2,3}$ coefficients are due to higher-order effects in
the
running of the parameters in $V_0$ (since the RGEs are solved numerically)
and are small relative to the linear term.)
The expected result is evident: $|b_1/a_1|=0.0025=1/400$, that is, $\wt V_1$ is
$Q$-independent to one-loop order (\ie, `flat' at $t=0$). It is important to
realize that the only test of the calculation is $b_1/a_1\ll 1$, since
$b_{2,3}$ are subject to two- and higher-loop corrections.

To study the relative importance of the various contributions to
${\rm STr}{\cal M}^4$ in Eq. \XX, in \fII\ we show the linear (in $t$)
contributions to $V_0$, and to $\wt V_1$ in two steps: (a) only
$t,\tilde t,b,\tilde b$, and (b) all contributions.
The corresponding coefficients are
\eqn\XXII{b_1^{(a)}=-0.0224,\qquad b_1^{(b)}=-0.0007,}
which give $|b_1^{(a,b)}/a_1|\sim 1/12,1/400$, showing an explicit
convergence to the correct result. It is clear that the ``dominant"
contributions to $\Delta V$ (\ie, $t,\tilde t,b,\tilde b$) are indeed the
largest ones. However, \fII\ shows that these may not be enough in applications
where the $Q$-independence of the one-loop potential over a several hundred GeV
range is essential.

Let us now comment on what happens if we allow the masses that enter into
$\Delta V(t,\phi)$ to run with $t$. To this end we rewrite $\Delta V(t,\phi)$
as follows
\eqn\XXIII{\Delta V(t,\phi)=-\coeff{1}{32\pi^2}t\,{\rm Str}{\cal M}^4(t,\phi)
+\coeff{1}{64\pi^2}{\rm STr}\,{\cal M}^4
\left(\ln{{\cal M}^2\over M_Z^2}-{3\over2}\right)(t,\phi).}
The leading-log piece ($\propto t$) displays the expected one-loop
$t$-dependence. However, the non-leading-log residue also has a linear term
in $t$ which is of two-loop order (\ie, $\propto1/(4\pi)^4$) and which
therefore spoils the vanishing of the linear term in the potential.
The magnitude of this effect is shown in \fII\ as the $\wt V'_1$ line, which
clearly deviates from the $\wt V_1\,({\rm full})$ line. Note that
the potential is formally $Q$-independent to one-loop no matter where the
masses in the supertrace are renormalized. However, only when these masses are
renormalized at $t=0$ do the two-loop subleading-log terms and the linear term
in the potential vanish.

In sum, we have shown that the subtracted one-loop effective potential
$\wt V_1$ is $Q$-independent to one-loop and should be used in calculations to
this order. Explicit running of the parameters in $\Delta V$ leads to a
residual
$Q$-dependence due to the introduction of spurious two-loop contributions
which affect the quadratic as well as the linear $t$-dependence of $\wt V_1$.
Note however that two- and higher-loop effects (\ie, the source for the
curvature of $\wt V_1$ in \fI{}) increase logarithmically with $Q/M_Z$ and
signal a progressive deterioration of the one-loop approximation. It is
therefore not advisable to use $\wt V_1$ to study effects over scales greater
than $1\TeV$. A renormalization-group-improved one-loop effective potential
\Sher\ (wherein all powers of $t$ are summed up) would be the proper tool for
this purpose.

\subsec{Numerical methods}
In this section we discuss the numerical methods and assumptions used to
explore the $Q$-(in)dependence of the tree-level and the one-loop Higgs
potentials as well as the one-loop minimization.
As we have demonstrated in the previous section, this procedure
involves the following basic steps: (i) the full scalar field-dependent
s/particle spectrum must be defined at a fixed scale, in order to calculate
$\Delta V$ (see appendix for details), (ii) a particular choice in the scalar
field space must be made in order to calculate $V_0,\wt V_1$, and (iii) the
parameters and fields in $V_0$, namely
$g_1(Q),g_2(Q),\mu(Q),B(Q),{m^2_{H_1}}(Q),{m^2_{H_2}}(Q),H_1(Q),H_2(Q)$ must be
RG evolved to the new scale in question. As we have shown, the issue of
$Q$-independence for $V_1$ thus becomes a numerical test as to whether the
implicit leading-log corrections to the tree-level parameters conspire to
cancel the explicit leading-log $Q$-dependence in $\Delta V$ (Eqn. \I).

For step (i), in order to solve for the `physical' theory at a fixed scale
$Q=M_Z$ and define $\Delta V$,
we begin by making a choice for the initial set of independent
parameters and integrate the RGEs to this scale. In order to calculate
$\Delta V$, the complete scalar field-dependent s/particle spectrum must be
defined. We then minimize the tree-level or one-loop potential to obtain a
consistent, complete set of parameters that represents the ground state of the
theory. One option would be to choose $\mu,B,m_0,m_{1/2},\lambda_t$ at $M_U$,
evolve to $Q=M_Z$ and then minimize $V_1(M_Z)$ or $V_0(M_Z)$ to determine
$v_1(M_Z),v_2(M_Z)$. From the experimental constraints for $M_W,M_Z$, the
original set of high-energy parameters are either allowed or ruled out. While
this approach involves the `direct' calculation of $v_1(M_Z),v_2(M_Z)$, the
original parameter space is at $M_U$, and the connection to low-energy physics
which include the constraints for $m_b,m_\tau$ must be done via an iterative
procedure.

Our calculational procedure for minimization of the one-loop potential is quite
different. For a given point in the five-dimensional parameter space
$m_t(m_t)$, $\tan\beta(M_Z)$, $m_{1/2}$, $m_0$, $A$ we begin by integrating the
RGEs for the gauge and Yukawa couplings up to $Q=M_U$ in order to specify the
complete set of boundary conditions at this scale. We then evolve back down to
$Q=M_Z$ but this time including the RGEs for the scalar masses as well.
The feasibility of this approach relies on the basic fact that the values of
$\mu,B$ decouple from the full set of RGEs. Thus, initial specification of
$\mu(M_U),B(M_U)$ is not required. Now armed with a set of
low-energy parameters (except for $\mu,B$, for which we must make an initial
guess), we can calculate the s/particle spectrum which enters into the
supertrace in $\Delta V$. We then solve for $\mu(M_Z)$ and $B(M_Z)$ via the
minimization conditions for the one-loop scalar potential. Specifically, we
numerically determine the values of $\mu$ and $B$ which solve the following
conditions
$${\left({\partial V\over \partial \phi_i}\right)_{\vev{\phi_3}=v_1,
\vev{\phi_7}=v_2,\vev{\phi_{1,2,4,5,6,8}}=0}=0,}$$
where $V=V(v_1,v_2,\mu,B,m_{1/2},m_0,A,m_t)$ is the scalar potential
(tree-level or one-loop), and the $\phi_i$ describe the eight real degrees of
freedom of the two Higgs doublets (in the notation of Ref. \GH). At tree-level
$V=V_0$ and the conditions above can be solved analytically for $\mu$ and $B$
(see Eqn. \Intii{}). For the one-loop potential, we employ a two-dimensional
Newton method which quickly locates the extremal values for $\mu,B$. We begin
by
making a guess for $\mu,B$ values, and then allow the system to
self-consistently relax to the extremum $\mu,B$ values.
Note that since we have demonstrated the one-loop $Q$-independence of
$\wt V_1(Q)$, this implies that if we were to minimize $\tilde V_1(Q)$ for
$Q\not=M_Z$, the values of $\mu(Q)$ and $B(Q)$ would just be the one-loop
RG-evolved $\mu(M_Z)$ and $B(M_Z)$ obtained by minimizing $\wt V_1(M_Z)$. We
can thus {\it derive} the boundary conditions for $\mu,B$ at any scale; we
conveniently choose $Q=M_Z$ such that
$M_Z^2={1\over2}(g_2^2(M_Z)+g'^2(M_Z))(v_1^2(M_Z)+v_2^2(M_Z))$
corresponds to the physical $Z$ mass.

In principle, there could be more than one point in the $\mu,B$ parameter space
which is consistent with the above conditions.
We have searched the $\mu,B$ parameter space via a Monte Carlo analysis
and find, however that in all cases considered, there is only {\it one} allowed
point for $\mu,B$. This simplifies the search of the parameter space
enormously.

As for step (ii), to evaluate the value of the potential itself, the discussion
above makes clear the natural choice for the configuration of the scalar field
space: the minimum of the potential at $Q=M_Z$. In general, it is not necessary
nor essential to choose the minimization configuration; due to the
non-negligible two-loop effects that we have described in Sec. 2.2, it is
likely that this initial choice in field space will no longer represent the
minimum for $Q>M_Z$ anyhow. Finally, regarding the tree-level parameters in
step (iii), we simply evolve the $Q$-dependent parameters
$g_1,g_2,\mu,B,{m_{H_1}}^2,{m_{H_2}}^2,v_1,v_2$ from $M_Z$ to variable $Q$,
using the standard one-loop RGEs. These running parameters contribute to the
implicit $Q$-dependence of $\wt V_1$ and appear solely in $V_0$.

\newsec{The fine-tuning problem}
Even though supersymmetric theories technically solve the gauge hierarchy
problem, it is well known that this problem may be re-introduced in a different
guise if the splitting of the supersymmetric multiplets exceeds
${\cal O}(1\TeV)$. A quantification of this `fine-tuning'
effect was proposed in Ref. \EENZ\ and later
elaborated on in Ref. \BG. The purpose of this section is to explore, in the
light of our own approach, the consequences of the proposed ``naturalness"
cut on the parameter space.
\subsec{General remarks}
The basic idea can be easily grasped by studying Eqn. \Intii{a}, which we
trivially rewrite as follows
\eqn\FTi{\coeff{1}{2}M^2_Z={m^2_{H_1}-m^2_{H_2}\tan^2\beta\over\tan^2\beta-1}
-\mu^2\equiv X_0m^2_0+X_{1/2}m^2_{1/2}-\mu^2.}
Clearly, for increasingly larger values of the dimensionful supersymmetry
breaking parameters, the renormalization-group evolved mass $m^2_{H_{1}}$
tends to increase, forcing $\mu^2$ to larger values as well. It can also
happen that the value of $m_t$ is fine-tuned such that $\mu$ remains small
even if $m_{1/2}$ and $m_0$ grow large. Therefore, the
measured value of $M_Z$ becomes the result of increasingly more ``fine-tuned"
cancellations among the supersymmetric masses.

Before proceeding to the specifics, we would like to point out a difference
between our approach and the standard one to the exploration of the parameter
space in this class of models. This difference is in the treatment of the
variables $\mu$ and $m_t$. We take $m_t$ as an input and obtain $\mu^2$
(which is related to $\mu_0=\mu(M_U)$ via RG-evolution) directly using
Eqn. \FTi. In the standard approach on the other hand, one gives
$\mu_0\leftrightarrow\mu$ and uses Eqn. \FTi\ as a {\it constraint} to
determine $m_t$ implicitly. The implied numerical inversion becomes harder
(\ie, more time-consuming) as the fine-tuning (\eg, $m_{1/2}$) grows. For
example, to obtain $\mu^2+{1\over2}M^2_Z$ with a relative accuracy of 0.001
(for $\mu=150\GeV$, $m_0=m_{1/2}$, $A=0$, $\tan\beta=5$) one needs to solve
for $m_t$ to the first/second/third decimal place for $m_{1/2}=150/300/600\GeV$
(one obtains: $m_t=120.8/97.45/91.061\GeV$). A variant of the standard
approach, wherein $m_t$ and $\mu$ are given as inputs and $m_0$ is solved
for using Eq. \FTi, has an analogous problem involving the determination of
$m_0$.
\subsec{The fine-tuning coefficients}
The usual definition of the fine-tuning parameters $c_i$ is given by
\refs{\EENZ,\BG}
\eqn\FTii{c_i=\left|{a_i\over M^2_Z}{\partial M^2_Z\over\partial a_i}\right|,
\quad a_i=\mu^2,m^2_t,m^2_{1/2},m^2_0,}
where the $a_i$ are the relevant parameters of the theory. It is then argued
that if $c_i<\Delta$, then cancellations among the parameters of at most
$\log(\Delta)$--orders of magnitude occur. The various expressions for the
$c_i$ scale with $(m_{1/2}/M_Z)^2$ and therefore we define scaled coefficients
$\hat c_i\equiv c_i/(m_{1/2}/M_Z)^2$, and obtain (with
$\wh M_Z\equiv M_Z/m_{1/2}$ and $\xi_0\equiv m_0/m_{1/2}$)
\eqna\FTiii
$$\eqalignno{
\hat c_\mu&=2\mu^2/m^2_{1/2}
                =2(X_{1/2}+X_0\xi^2_0-\coeff{1}{2}\wh M^2_Z),&\FTiii a\cr
\hat c_t&=2m^2_t\left|{\partial X_0\over\partial m^2_t}\xi^2_0
                +{\partial X_{1/2}\over\partial m^2_t}\right|,&\FTiii b\cr
\hat c_{1/2}&=2\left|X_{1/2}\right|,&\FTiii c\cr
\hat c_0&=2\left|X_0\right|\xi^2_0.&\FTiii d\cr}$$
The important point is that since the $\hat c_i$ scale with $m^2_{1/2}$, then
an upper bound on $m_{1/2}$ (for a given $\xi_0$) results for a given choice of
$\Delta$, \ie, $m_{1/2}<m^{max}_{1/2}(\xi_0,\Delta)$, and these bounds scale
with $\sqrt{\Delta}$.\foot{There is of course also an upper bound on
$m_0=\xi_0 m_{1/2}<\xi_0 m^{max}_{1/2}(\xi_0,\Delta)$ (unless
$X_0=\partial X_0/\partial m^2_t=0$; see below).}

To facilitate the subsequent discussion, we now give analytical expressions
for the quantities $X_0,X_{1/2}$ \refs{\BG,\Japspd} which are valid in the
limit of vanishing $\lambda_b,\lambda_\tau$, or equivalently for not too
large $\tan\beta$ (\ie, $\tan\beta<8$ \Japspd)
\eqna\FTiv
$$\eqalignno{
X_0&=-1+\coeff{3}{2}{\tan^2\beta+1\over\tan^2\beta-1}
                                \left({m_t\over m^0_t}\right)^2,&\FTiv a\cr
X_{1/2}&=-K_l+{\tan^2\beta+1\over\tan^2\beta-1}\left({m_t\over m^0_t}\right)^2
\left\{a+\coeff{1}{2}(b-\xi_A)^2-\coeff{1}{2}(b-\xi_A)^2
\left({m_t\over m^0_t}\right)^2(1+\tan^{-2}\beta)\right\},\cr
                                                        &&\FTiv b\cr}$$
where $\xi_A\equiv A/m_{1/2}$, $m^0_t=192\GeV$, and $K_l,a,b$ are some
(positive) numerical coefficients.

The simplest fine-tuning parameter is $\hat c_\mu$, which basically imposes
an upper bound on $\mu$,
\eqn\FTv{|\mu|<\mu^{max}=\sqrt{\Delta\over2}\,M_Z.}
However, since in our approach $\mu$ is a derived quantity, the constraints
on the parameters of the theory are less transparent. In \fIII{a} we show
$\hat c_\mu$ (calculated exactly and for $\tan\beta=5$) as a function of
$\xi_0$ (for $\xi_A=0$)
and for three values of $m_t$. From Eq. \FTiv{a} we see that $X_0$ has a
zero at $m_t\approx151\GeV$ (for $\tan\beta=5$) and therefore $\hat c_\mu$
will be quadratic in $\xi_0$ but with negative (positive) curvature for
$m_t<151\GeV$ ($m_t>151\GeV$), as observed in the figure. The magnitude of the
curvature grows with $m_t$, also as anticipated. In \fIII{b} we show the
case where $\xi_A=\xi_0$, which has a similar behavior, although the zero of
$X_0$ is not the relevant turning point anymore since $X_{1/2}$ also depends on
$\xi_0$ through $\xi_A$.

{}From the expression for $\hat c_\mu$ one can see that if $X_0=0$, then even
though $m_{1/2}$ would still need to be bounded above, $m_0$ would not.
Analogously, if $X_{1/2}=0$, then $m_{1/2}$ could grow indefinitely
without affecting $c_\mu$. These peculiar points in parameter space
\BG\ are isolated and are not stable in perturbation theory. Moreover, in
the second instance, they correspond to small values of $\mu$ which are
excluded on phenomenological grounds. In the figures, the first case is seen
to occur in the $m_t=150\GeV$ curve in \fIII{a}, whereas the second case is
approached by the $\xi_0=\xi_A\approx7$ point on the $m_t=145\GeV$ curve
in \fIII{b}.

The $\hat c_t$ fine-tuning coefficient is shown in \fIV\ in analogy to
\fIII. Large values of this coefficient correspond to instances in which $m_t$
is fine-tuned to give small values of $\mu$ in Eq. \FTi, even though $m_{1/2}$
and $m_0$ grow large. This coefficient gives qualitatively similar constraints
since it is also quadratic in $\xi_0$. From the approximate expression for
$X_0$ one can
easily show that $m^2_t\partial X_0/\partial m^2_t=X_0+1>0$, and therefore
we expect $\hat c_t$ to grow with $\xi^2_0$ for all allowed values of $m_t$,
and also $\hat c_t>\hat c_\mu$. These facts are evident in the figures.

The effect of the $\hat c_{1/2}$ coefficient can be studied from $\hat c_\mu$
by setting $\xi_0=0$ ({\it c.f.} Eqs. \FTiii{a,c}). As seen from \fIII\ (for
$\xi_0=0$), $\hat c_{1/2}$ depends very weakly on $m_t$ and therefore gives a
direct
upper bound on $m_{1/2}$, independently of $\xi_0,\xi_A$, although this bound
is not as strong as the ones obtained from $\hat c_{\mu},\hat c_t$. Finally,
the $\hat c_0$ coefficient has the same $\xi_0$ dependence as does $\hat
c_\mu$,
although its magnitude is shifted down by $X_{1/2}$.
\subsec{Some sample bounds}
It is not our intention to quote concrete upper bounds on the various
parameters
in the model, since these would depend on the chosen value for the cutoff
$\Delta$ (although these bounds would scale with $\sqrt{\Delta}$). For example,
from Eq. \FTv\ taking $\Delta=10$ (as done in previous analyses \refs{\BG,\RR})
gives $|\mu|\lsim200\GeV$, and $|\mu|/M_Z\lsim2.2$. This bound appears
unnecessarily stringent. A more reasonable cutoff of $|\mu|\lsim450\,(650)\GeV$
is obtained for $\Delta=50\,(100)$. It is also not clear to us whether the
above
definition of the $c_i$ is the only possible one, or rather whether alternative
definitions would yield quantitatively similar results (qualitatively they must
all agree).

Our purpose here is to determine whether the values of $m_{1/2},\xi_0,\xi_A$
which we will examine later give values of the $c_i$ below a `reasonable'
cutoff of say $\Delta=50-100$. In general we have
$m_{1/2}<\sqrt{\Delta/\hat c_{\mu,t}}\,M_Z$. From the figures it is clear that
for $\xi_0\lsim1$ one gets $\hat c_{\mu,t}\approx3-6$, and therefore
$m_{1/2}\lsim370-265\,(525-370)$ for $\Delta=50\,(100)$. These estimates hold
for $\tan\beta=5$ and decrease with decreasing $\tan\beta$: for $\tan\beta=2$,
$\hat c_\mu(\xi_0\lsim1)\approx9$, $m_{1/2}\lsim220\,(310)\GeV$ for
$\Delta=50\,(100)$. Larger values of $\xi_0$ strengthen these bounds
rapidly.\foot{Note that larger values of $\xi_0$ are fine as long as $m_{1/2}$
is low enough, \eg, for $m_{1/2}\lsim100\GeV$, $c_t<50\,(100)$ for
$\xi_0\lsim4\,(6)$.}
We thus see that if we restrict $m_{1/2}$ to $m_{1/2}\lsim400\GeV$ or
equivalently $m_{\tilde g}\lsim1\TeV$ (and take $\xi_0\lsim1$) then the
resulting class of models will presumably remain in the reasonable fine-tuning
regime. Since the squark masses can be approximated by $m^2_{\tilde q}=
m^2_{1/2}(c+\xi^2_0)$, with $c\approx6$, we also see that $m_{\tilde q}\lsim
1\TeV$.

Even though these fine-tuning ``bounds" on $m_{1/2}$ and $\xi_0$ are not very
precise and in fact do not necessarily have to hold in the real world, it is
useful to have an estimate of where things start to become ``un-natural".

\newsec{The models and their motivation}
For the purposes of this paper we will consider two $SU(3)\times SU(2)\times
U(1)$ supersymmetric models.  The first is the supersymmetric Standard
Model (SSM) with the minimal three generations and two Higgs doublets of matter
representations, and which is assumed to unify into a larger gauge group at a
unification mass of $M_U\approx10^{16}\GeV$.
The second is the String-Inspired Standard Model (SISM) \SISM\ with additional
vector-like $Q$ and $D^c$ matter representations with masses set to obtain
$\sin^2\theta_w=0.233$ and a string unification scale of $M_U=10^{18}\GeV$.

However, it is widely believed that
there must be more than this to nature, in particular grand-unification
and string unification.  Surprisingly, many grand unified and string
models can reduce (below the unification scale) to the SSM or SISM with a few
minor alterations.  Thus these simpler models can give a good first
approximation for the low-energy predictions of more realistic models, and
provide the groundwork upon which the extra details of more complicated models
may be added.

Take for example minimal supersymmetric $SU(5)$.  After the GUT is broken,
the light content of the model is exactly that of the SSM.  Probably the
most important difference are the dimension-five proton-decay-mediating
operators resulting from integrating out heavy GUT Higgs triplet fields of
mass $M_H$. The presence of these operators imposes strong constraints on the
parameters of the model \refs{\CEN,\Japspd,\AN}. Indeed, in Ref. \AN\ it is
shown that the current experimental lower bound on the decay mode
$p\to\bar\nu K^+$ can be transformed into an upper bound on a quantity $P$
(called $B$ in Ref. \AN) which encompasses
the sparticle mass dependences of the decay width ($\tau\propto1/P^2$).
These authors find $P<(103\pm15)(M_H/M_U)$.
The function $P$ is complicated but it is argued to be reduced by small
gluino masses and large scalar masses, \ie, by large values of $\xi_0$.
It also has the following explicit dependence on $\tan\beta$,
$P\propto(1+\tan^2\beta)/\tan\beta$, which then reduces $P$ for small values of
$\tan\beta$. How large does $\xi_0$ need to be to obtain acceptable proton
decay rates? In Table I we show the values of the fine-tuning coefficients
$c_\mu$ and $c_t$ which
follow from the choices made in fig. 2 of Ref. \AN\ for the model parameters
($\tan\beta=1.73$, $m_t=125\GeV$, $\mu>0$, and $A_t=-0.6m_0$). The minimum
values of $\xi_0$ are those required to obtain a value of $P$ equal to its
experimental upper bound and with the assumption $M_H=3M_U$. We see that the
fine-tuning coefficient $c_t$ exceeds $\Delta=100$ in all cases and therefore
compatibility with proton decay experiments drives the minimal model into the
fine-tuned regime. This was also observed in Ref. \AN. Our point here is to be
more quantitative in the light of our own studies of the fine-tuning
constraint. Note that increasing the value of $\tan\beta$ increases $P$ and
therefore makes the experimental bound harder to satisfy. Note also that
had we chosen $M_H=M_U$ instead, then the minimum $\xi_0$ values get
increased significantly (refer to Table I): for $m_{1/2}=74\,(122)\GeV$,
$\xi^{min}_0$ goes from $8.1\,(6.5)$ to $16.2\,(11.5)$ and $c_t$ from
$113\,(204)$ to $432\,(603)$.

Clearly, the minimal SU(5) supergravity GUT is a rather constrained model,
which may not survive proton decay bounds if the fine-tuning
constraints hold at face value.\foot{The minimal model with a Higgs singlet
must also contend with the doublet-triplet splitting problem \dt. The missing
partner mechanism \MNTY\ solves this potential problem at the expense of
introducing new large GUT massive representations. Nevertheless, the light
content of the model is still the SSM.} Therefore, the SSM is probably best
thought of as the low-energy limit of a unified model where the strict proton
decay constraints are naturally avoided.
One strategy to remedy the ailments of the minimal $SU(5)$ model is to change
the gauge group to flipped $SU(5)$ \Barr.  After GUT breaking, the resulting
effective theory is the SSM with the Higgs mixing term $\mu\bar h h$ provided
elegantly by the superpotential terms $\lambda_7\bar h h\phi+\lambda_8\phi^3$
involving a singlet field \AEHN.  This is qualitatively the
same as the usual Higgs mixing term once the singlet gets a vev.
However, there is a crucial difference relative to the minimal $SU(5)$ model
since now the Higgs triplet mixing term ($\propto\vev{\phi}$) is naturally
of ${\cal O}(M_W)$ and leads to a suppression of the dangerous dimension-five
proton decay operators of ${\cal O}(M_W/M_U)$ relative to the minimal $SU(5)$
case, thus making these operators innocuous \aspects. Quantitatively, there are
differences in the two models, though the dimension of the solution space is
the same.  The tree-level case may be solved by noting that the dimensionful
parameters in the tree-level potential scale with $m_{1/2}$, allowing one to
minimize a dimensionless potential and then fix $m_{1/2}$ from the measured
value of $M_Z$ \aspects.  This approach does not work in the one-loop case
because the spectrum does not scale with $m_{1/2}$, and $\lambda_7$ and
$\lambda_8$, unlike $\mu$ and $B$, feed into the other RGEs.  Solving the
one-loop problem requires a search over parameters at the unification scale for
the subspace which gives the correct $M_Z$, the needle in the haystack
approach.
In string-derived flipped models, the role of the singlet
vev is played by a hidden sector condensate in a non-renormalizable
superpotential term which effectively reproduces the usual Higgs mixing
term in the SSM \decisive.

Another possibility is that there is no GUT.  The requirement of gauge coupling
unification when there is no GUT is unmotivated in field
theory but completely natural in string theory, and models exist in
which the string directly gives rise to the Standard Model gauge group
\refs{\SMCY,\SMOrb,\SMFFF}. String models predict a unification scale of about
$10^{18}\GeV$ \refs{\Kap,\Lacaze} and usually contain several extra vector-like
matter representations.  This motivates
the SISM, in which the extra vector masses are choosen so that gauge
unification scale gives the required $10^{18}\GeV$ result \SISM.  Among the
many possible sets of extra vector representations which might arise
from the string and realize the correct unification scale, we have choosen
the unique minimal set for the SISM to analyze in this work. These are an
extra vector-like quark doublet with mass $m_Q\approx3\times10^{12}\GeV$ and
an extra vector-like charge $-1/3$ quark singlet with mass $m_{D^c}\approx
3\times10^5\GeV$ \SISM.

One concern is that the effective SSM or SISM from more complicated
models might not result in universal soft supersymmetry breaking.  If
the minimal $SU(5)$ model had universal soft supersymmetry breaking at
a scale much larger than the GUT scale, differences in the supersymmetry
breaking masses between the \r{10} and \rb{5} representations would develop.

In the flipped model, however, it can be shown that the onset of
supersymmetry breaking, the $SU(5)\times U(1)$ and the
$SU(5)\times U(1)\to SU(3)\times SU(2)\times U(1)$
scales must be within about an order of magnitude of each
other \aspects, and that the modifications of universal soft supersymmetry
breaking in the resulting SSM are therefore small, in accordance also with the
natural absence of dangerous flavor-changing neutral currents (FCNC) which
are endemic in supersymmetric theories \EN. Coupling constant unification
and the experimental bounds on the low-energy gauge couplings require
that the two gauge unification scales in the flipped model be almost
identical \price.  The GUT breaking is along a flat direction in the absence
of supersymmetry breaking of the potential which develops a minimum upon
the introduction of soft supersymmetry breaking terms.  Dynamical
calculations result in a GUT scale which is just slightly below the scale
where supersymmetry breaking is introduced \aspects.

In summary, the simple SSM and SISM models not only have the
advantage of ignoring the complicated and model-dependent details of
more realistic GUT and string models, but also capture the essence of
many of these more realistic models as these extra details, in many cases,
are irrelevent or give small corrections to the results of the simpler
models.

\newsec{The allowed parameter space}
\subsec{Constraints on the parameter space}
We now present the several consistency and experimental constraints on this
class of models and later discuss how they restrict the allowed parameter
space.\foot{For reference, the values of the measured
parameters which have been used are: $\alpha_3=0.113$, $m_b=4.9\GeV$,
$\alpha^{-1}_{em}=127.9$, $M_Z=91.2\GeV$, and $m_\tau=1.784\GeV$. The allowed
regions in parameter space have a small sensitivity to $\delta\alpha_3=0.004$.}
\smallskip
\noindent {1. Consistency constraints}\nextline
(i) Perturbative unification: the values of the $t,b$, and $\tau$ Yukawa
couplings should remain in the perturbative regime (and certainly finite) all
the way up to $M_U$. Tree-level partial wave unitarity is violated if these
couplings exceed $\lambda\approx5$ at any scale \DL. We apply these relations
at $M_U$ where they are most constraining. At low energies these upper bounds
on the Yukawa couplings get transmuted into upper bounds on $m_t$ and
$\tan\beta$ as follows,
\eqn\Ai{m_t=174\lambda_t\sin\beta<174\lambda^{max}_t/
\sqrt{1+1/\tan^2\beta}\lsim190\pm1\GeV,}
for $\alpha_3=0.113\pm0.004$, and
\eqn\Aii{m_b=174\lambda_b/\sqrt{1+\tan^2\beta}
\lsim174\lambda_b^{max}/\sqrt{1+\tan^2\beta}\Rightarrow\tan\beta\lsim47\pm2}
for $m_b(m_b)=4.9\mp0.1\GeV$. Since dynamically one gets $\tan\beta>1$ (see
below) and experimentally $m_t>90\GeV$, we get a completely bounded region in
the $(m_t,\tan\beta)$ plane.
\smallskip
\noindent (ii) Electroweak breaking: as discussed in Sec. 2.3, we solve for
$\mu$ and $B$ using the one-loop effective potential. This implies \GRZ\
that the set of subsidiary conditions which are usually imposed to obtain a
good minimum of the tree-level potential (\ie, boundedness, stability, and
avoidance of electric charge and color breaking minima) are automatically
satisfied by the one-loop potential and do not need to be (and have not been)
imposed. It is however necessary to demand boundedness of the potential at the
unification scale, \ie,
\eqn\Aiii{{\cal B}=m^2_{H_1}+m^2_{H_2}+2\mu^2+2B\mu\to
2(m^2_0+\mu^2_0+B_0\mu_0)
>0.}
We also demand that all squared squark, slepton, and Higgs masses be positive.
In particular, this must hold for the pseudoscalar Higgs mass, $m^2_A>0$.
At tree-level $m^2_A=-2B\mu/\sin2\beta$ and therefore $(B\mu)_{tree}<0$.
However, this does not necessarily hold at one-loop since there are additional
contributions to $m^2_A$ which allow both signs of $(B\mu)_{loop}$.
\smallskip
\noindent (iii) Cosmology: astrophysical considerations indicate that the
lightest supersymmetric particle (LSP) must be neutral and colorless \EHNOS.
This leaves two candidates: the sneutrino and the lightest neutralino. As
discussed in Ref. \KLNPY, in most of the parameter space it is the lightest
neutralino which is the LSP, and is sensible to neglect the small regions of
parameter space where the sneutrino is the LSP (\ie, $m_{\tilde\nu}\approx
42-46\GeV$\foot{If the sneutrinos make up the dark matter in the galactice
halo,
this mass range is eliminated by LEP measurements \ENRS.}). In what follows we
exclude points in parameter space where the lightest neutralino is not the LSP.
\smallskip
\noindent (iv) Naturalness: as discussed in Sec. 3 to avoid re-introducing
the fine-tuning problem, we require $m_{1/2}\lsim400\GeV$ (\ie, $m_{\tilde g}
\lsim1\TeV$) and $\xi_0\lsim1$ (or more properly, $c_{\mu,t}\lsim100$).
\medskip
\noindent {2. Experimental constraints}\nextline
We impose the following cuts on the sparticle masses and/or couplings:\nextline
\noindent(i) The LEP lower bound on the chargino mass
$m_{\chi^\pm}>45\GeV$ \chargino.\nextline
\noindent(ii) The CDF lower bounds on the gluino ($m_{\tilde g}>150\GeV$)
and the squark ($m_{\tilde q}>100 \GeV$) masses \CDF. These bounds are
actually correlated and are subject to numerous assumptions. Fortunately,
they are satisfied automatically in this class of models once the other
constraints are imposed.\nextline
\noindent(iii) The LEP lower bound on the charged slepton mass
$m_{\tilde l}>43\GeV$ \PDG.\nextline
\noindent(iv) The CDF lower bound on the top quark mass $m_t\gsim90\GeV$ \top.
We do not expect a significant weakening of this SM bound due to potential
non-SM top decay processes, since these are rather suppressed (\eg,
$t\to bH^+$, $m_{H^+}>M_W$, $t\to\tilde t_1\chi^0_1$,
$m_{\tilde t_1}+m_{\chi^0_1}>72\GeV$ in this model).\nextline
\noindent(v) The contributions to the invisible and new $Z$ widths from
$Z$ decay into neutralino pairs, \ie,
$\Gamma(Z\to \chi^0_1\chi^0_1)<\Gamma^{inv}<18\MeV$ and
$\Gamma(Z\to \chi^0_i\chi^0_j)<\Gamma^{new}<28\MeV$ ($i=j=1$
excluded) at 95\% C.L. \widths.\nextline
\noindent(vi) The contribution to the invisible $Z$ width from
$Z\to \tilde\nu\tilde{\bar\nu}$ for $m_{\tilde\nu}\le{1\over2}M_Z$, \ie,
$\Gamma(Z\to \tilde\nu\tilde{\bar\nu})<18\MeV$, since we expect the invisible
decay mode $\tilde\nu\to\nu\chi^0_1$ to dominate in this $m_{\tilde\nu}$
range. For three degenerate sneutrinos, as is the case in the two models we
consider, we get
$m_{\tilde\nu}\gsim42\GeV$. Realistically this bound will be further pushed
towards ${1\over2}M_Z$ since the $\Gamma(Z\to \chi^0_1\chi^0_1)$
also contributes to $\Gamma^{inv}$ in the same region of parameter space,
thus possibly closing up the window for sneutrino LSP.\nextline
\noindent (vii) The experimental constraints on the lighter
Higgs bosons $h$ and $A$, as follows: we require $\Gamma(Z\rightarrow h Z^*
\rightarrow h \mu^+\mu^-)/\Gamma(Z\rightarrow \mu^+\mu^-)<5\times 10^{-5}$,
and $\Gamma(Z\rightarrow h A)/\Gamma(Z\rightarrow \mu^+\mu^-)<0.11$; these
values were obtained from a graphical fit to the experimental results
\refs{\ALHiggs,\widths}. The Higgs boson masses have been calculated to
one-loop accuracy as described in Ref. \Higgs.

\subsec{Anatomy of the allowed regions in $(m_t,\tan\beta)$ space}
The consistency constraints described above allow us to obtain a bounded
region in $(m_t,\tan\beta)$ space for all values of $m_{1/2},\xi_0,\xi_A$.
These areas are further constrained by the phenomenological cuts given
above. It is however more illuminating to understand the shape of the
bounded region prior to these cuts. This is particularly simple in the
tree-level case where $\mu$ and $B$ are obtained directly from Eqs. \Intii{}.
In this case one has to demand two subsidiary conditions to ensure a good
symmetry breaking minimum, namely that the tree-level potential be bounded
from below
\eqn\Aiv{{\cal B}=m^2_{H_1}+m^2_{H_2}+2\mu^2+2B\mu>0,}
and that it possesses a minimum away from the origin of field space,
\eqn\Av{{\cal S}=(m^2_{H_1}+\mu^2)(m^2_{H_2}+\mu^2)-B^2\mu^2<0.}
In the one-loop case $\mu$ and $B$ need to be solved for numerically and
these conditions are enforced automatically \GRZ. But then,
when is an $(m_t,\tan\beta)$ point not allowed at one-loop? This happens
when the search for a $\mu,B$ pair fails because $B$ is driven to very large
values (\ie, $\mu\to0$) or when an otherwise acceptable pair nevertheless gives
$m^2_A<0$.

To illustrate our remarks, in \fV{a} and \fV{b} we show the resulting allowed
regions for $\xi_0=0,1$ and $\xi_A=0$ for $m_{1/2}=150,250\GeV$ and both signs
of $\mu$ for the SSM {\it prior} to the application of any phenomenological
cuts. (The following discussion applies qualitatively to the SISM case also.)
The solid (dashed) boundaries are those calculated using the tree-level
(one-loop) potential. The dotted lines are one-loop contours of $\mu$ (for
$\mu>0$ boundaries) and $B$ (for $\mu<0$ boundaries). The tree-level and
one-loop boundaries have five distinct portions. These are most easily
understood in tree-level approximation as follows:
\smallskip
\item{(i)}The top boundary: this is the positively sloped line restricting
the values of $\tan\beta$. For points above this line ${\cal B}<0$. For
points on or below this line we can use Eqs. \Intii{} to obtain
\eqn\Avi{{\cal B}=(1-\sin2\beta)\left[{\tan^2\beta+1\over\tan^2\beta-1}
(m^2_{H_1}-m^2_{H_2})-M^2_Z\right].}
Clearly since $m_{H_{1,2}}$ scale with $m_{1/2}$, for $m_{1/2}\gg M_Z$
an asymptotic state is obtained in which the explicit $M_Z$ dependence
becomes irrelevant, whereas any finite value of $m_{1/2}$ helps drive
${\cal B}$ to negative values and therefore more restrictive top boundaries
result, as the figures show. Note though that $m_{1/2}=250\GeV$
is already quite asymptotic. For what values of $m_t,\tan\beta$ is this
constraint relevant (or where in the plane does this line lie)? Consider
the following RGE
\eqn\Avii{{d\over dt}(m^2_{H_1}-m^2_{H_2})=\coeff{1}{8\pi^2}(3\lambda^2_bF_b
-3\lambda^2_tF_t),}
where $F_{b,t}$ are linear combinations of squared supersymmetry breaking
masses. At $M_U$, $m^2_{H_1}=m^2_{H_2}$ and $F_b=F_t$ and therefore for
$\lambda_b<\lambda_t$, $m^2_{H_1}-m^2_{H_2}$ grows as $t$ decreases. One needs
$\lambda_b\approx\lambda_t$ to turn this around and have $m^2_{H_1}-m^2_{H_2}
<0$ at $M_Z$, otherwise ${\cal B}>0$ automatically. From Eqs. \Ai\ and \Aii\
we get for $\tan\beta\gg1$, $\lambda_t\approx m_t/174$ and
$\lambda_b\approx \tan\beta\,m_b(M_Z)/174$, where $m_b(M_Z)=3.77\GeV$ (for
$m_b(m_b)=4.9\GeV$).
Therefore the ${\cal B}\approx0$ line occurs for $\tan\beta\approx m_t/3.77$,
as observed in the figures. This result is independent of $\xi_0,\xi_A$.
For later reference, from Eqs. \Intii{b} and \Aiv\ it is easy to see that
$m^2_A=-B\mu/\sin2\beta\approx0$ is equivalent to ${\cal B}\approx0$,
\ie, on the top boundary $m^2_A=0$.
\item{(ii)}The upper corner: the perturbative unification constraint in \Aii\
cuts off the growth of the top boundary. The rounded portion at the top and
towards the side of increasing values of $m_t$ results from the strengthening
of the perturbative cut on $\tan\beta$ due to the increasing value of
$\lambda_t$ \DL.
\item{(iii)}The right boundary: this originates from the perturbative cut
on $m_t$ in Eq. \Ai. The curved portion for small $\tan\beta$ also follows
from \Ai\ for the corresponding values of $\tan\beta$, \eg, we get
$m^{max}_t=135,171,181\GeV$ for $\tan\beta=1,2,3$.
\item{(iv)}The bottom boundary: From Eq. \Intii{a} we see that if $\tan\beta<1$
then $\mu^2<0$ since $m^2_{H_1}-m^2_{H_2}>0$ for $\lambda_t\gg\lambda_b$,
as is true for $\tan\beta\lsim1$ (see discussion following Eq. \Avii).
Therefore $\tan\beta\ge1$ always.\foot{In the $\tan\beta=1$ case at tree-level
one gets $m^2_{H_1}=m^2_{H_2}$, $B\mu=-(m^2_{H_1}+\mu^2)$, and therefore
${\cal B}=0$, ${\cal S}=0$, \ie, a rather special scenario which also gives
$m_h=0$. One-loop corrections to $V_0$ are most important in this case \KLNQ.}
\item{(v)}The left boundary:
\itemitem{a.} $\mu>0$: points to the left of this line have $\mu^2<0$
at tree-level. From Eq. \FTi\ we have
\eqn\Aviii{\mu^2=m^2_{1/2}(X_{1/2}+X_0\xi^2_0-\coeff{1}{2}\wh M^2_Z),}
with $X_{0,1/2}$ given in Eqs. \FTiv{} in the limiting case of small
$\tan\beta$. For the values of $m_t$ where this boundary occurs, $X_0$ is
negative (it turns positive for $m_t=97,121,140\GeV$ for $\tan\beta=1.5,2,3$)
and therefore increasing $\xi_0$ tends to drive $\mu^2$ to negative values,
unless this is compensated by an increase in $X_{1/2}$, \ie, by increasing
$m_t$. Therefore raising $\xi_0$ shifts the left boundary to the right, as
shown in the figures. The effect of $\xi_A$ is less pronounced. For fixed
$\xi_0$, Eq. \FTiv{b} shows that $\xi_A>0$ ($\xi_A<0$) decreases (increases)
$X_{1/2}$ since $b>0$. That is, a larger (smaller) value of $m_t$ is needed
to obtain the same $\mu^2$ and therefore $\xi_A>0$ ($\xi_A<0$) shifts the
left boundary to the right (left).
\itemitem{b.}$\mu<0$: for $\xi_0\gsim0.1$ the same remarks as the $\mu>0$
case apply. For $\xi_0\lsim0.1$ a peculiarity occurs due to the boundedness
at the unification scale constraint in Eq. \Aiii. In this case
${\cal B}_0\approx2(\mu^2_0+B_0\mu_0)$ and for $\mu_0<0$ (and therefore
$\mu<0$) too small values of $\mu$ can give ${\cal B}_0<0$. This results in
the peculiar shape of the $\mu<0$ boundaries in \fV{a}.
\smallskip
As seen from \fV{a} and \fV{b}, the one-loop minimization yields very similar
boundary configurations. In fact, only the left and (to a lesser extent the)
top boundaries show any shift relative to their tree-level counterparts, as
easily inferred from the above discussion on the origins of the various
portions of the boundaries. Just outside the left boundary one finds that
$B$ is driven to very large values (\ie, $\mu\to0$), and just outside the top
boundary $m^2_A<0$. These are precisely the same conditions defining these
boundaries at tree-level.

The shape of the $\mu$ contours can be deduced from Eq. \Aviii, at least
at tree-level and for small $\tan\beta$. For fixed $m_{1/2}$ and $\xi_0$,
the value of $\mu$ becomes independent of $\tan\beta$ for large $\tan\beta$
(see Eq. \FTiv{}) and increases with $m_t$ (at least as long as $X_{1/2}$
dominates over $X_0\xi^2_0$). This behavior is indeed realized numerically.
Surprisingly, it is found to persist for large values of $\tan\beta$ {\it and}
at the one-loop level, as the figures show. In fact, for not too small values
of $\tan\beta$, the tree-level contours have a `shadow' one-loop contour
running on top and viceversa. The relation between these contours is shown
in Table II for the typical case of $\xi_0=\xi_A=0$ in \fV{a}. One can see
that the shift in $\mu$ is largest for small $\mu$ and then asymptotes to a
constant value which grows with $m_{1/2}$. (Similar results are obtained for
the $\xi_0=1,\xi_A=0$ case in \fV{b}.)

In fact, the shift is {\it maximum} for $(m_t,\tan\beta)$ points on the
one-loop
left boundary, since there $\mu_{loop}=0$ by definition. This means that on
that
boundary the relative change in $\mu$ is $100\%$, and decreases as one
moves away towards larger values of $m_t$. The fact that the one-loop corrected
potential induces a $100\%$ shift on $\mu$ indicates that the underlying
perturbative approach to the determination of $\mu$ breaks down in this region
of parameter space. Fortunately, these points (where $\mu_{loop}\approx0$) are
ruled out on phenomenological grounds: (i) for $\mu=0$ the LSP is a massless
Higgsino state which gives $\Gamma(Z\to\chi\chi)=\Gamma(Z\to\nu\bar\nu)
[(\tan^2\beta-1)/(\tan^2\beta+1)]^2<18\MeV$ for $\tan\beta<1.4$ only;
(ii) for larger values of $\mu$ (but still close to zero) the lightest chargino
falls below the LEP lower bound. Therefore one does not have to worry too
much about the possible effect of two- and higher-loop contributions to $\mu$.

The figures also show (although perhaps not very clearly)
that the values of $\mu$ scale with $m_{1/2}$. At tree-level, the
$-|\mu_{tree}|$ and $|\mu_{tree}|$ contours overlap in $(m_t,\tan\beta)$ space.
Even though this is not supposed to be necessarily the case for their one-loop
counterparts, our results indicate that this is very approximately the case
too. Therefore, the $|\mu|$ contours on the $\mu>0$ boundaries in \fV{} can be
mapped onto the $\mu<0$ boundaries as $-|\mu|$ contours on the same positions
in the $(m_t,\tan\beta)$ plane.

A similar analysis for the $B$ contours (shown at one-loop on the $\mu<0$
boundaries in \fV{}) reveals that `shadow' contours exist only for the larger
values of $B$. Note also that contrary to the tree-level case where $B_{tree}$
vanishes at the top boundary, $B_{loop}$ does not. This is because on the
top boundary $m^2_A=0$ and $(m^2_A)_{tree}=-(B\mu)_{tree}/\sin2\beta=0$
for $B_{tree}=0$, whereas $(m^2_A)_{loop}=-(B\mu)_{loop}/\sin2\beta+\Delta
m^2_A
=0$ for $B_{loop}\not=0$.
\subsec{Effects of phenomenological cuts on the parameter space}
We have explored the five-dimensional parameter space by determining the
slices in $(m_t,\tan\beta)$ which are allowed for both signs of $\mu$,
$m_{1/2}=150,250\GeV$, and several choices of $\xi_0,\xi_A$, which respect
the fine-tuning constraints. Here we have imposed all the phenomenological cuts
discussed above (see \fVI{}). (However, for clarity of presentation, in the
figures we do not restrict the values of $m_t$.)
These discrete choices for $m_{1/2},\xi_0,\xi_A$ represent a good sample of the
full space and are enough to infer the shape of the $(m_t,\tan\beta)$ region
for
other values of the parameters.

The phenomenological cuts on the parameter space have various degrees of
effectiveness. For $\xi_0\lsim0.1$ the neutral LSP cut is most constraining.
This is because for increasing $m_{1/2}$ the LSP tends to become the lightest
$\tilde\tau$ state. Indeed, since we have not neglected $\lambda_\tau$ in the
calculation, the off-diagonal elements in the $\tilde\tau$ squared mass matrix
($\propto m_\tau(A_\tau+\mu\tan\beta)$) grow with $\mu$ (which grows with
$m_t$) and push down the $\tilde\tau_1$ mass.\foot{Had we set $\lambda_\tau=0$,
a similar effect would still be present with $\tilde e_R$ instead of
$\tilde\tau_1$, although with no significant $m_t$ dependence.} To avoid a
charged LSP the range of $m_t$ has to be cut off as shown in \fVI{a}.
In particular, \fVI{a} (solid line) should be compared with \fV{a} where no
phenomenological cuts were imposed; the severity of the cut is stunning.
The value of $\xi_A$ has a small effect on the magnitude of this cut as far
as the direct effect of $A_\tau$ is concerned (for large values of
$\tan\beta$).
However, $\xi_A$ influences the value of $\mu$ also (see Eqs. \Aviii\ and
\FTiv{b}). In fact, $\xi_A>0$ ($\xi_A<0$) decreases (increases) $X_{1/2}$
and therefore $\mu$ (for fixed $\xi_0$). Thus $\xi_A>0$ ($\xi_A<0$) weakens
(strengthens) the LSP cut by shifting $m_{\tilde\tau_1}$ upwards (downwards),
as is evident in \fVI{} (\ie, compare the dashed ($\xi_A=-1$), solid
($\xi_A=0$), and dotted ($\xi_A=1$) boundaries).
Note that the $\mu<0$ figures suffer from a less
effective LSP cut. This is because for the same $|\mu|$ and $m_{1/2}$ values
(and low $\tan\beta$), $m_\chi$ is lower for $\mu<0$.
The LSP cut becomes ineffective for $\xi_0\gsim0.1$ since by then the sleptons
get an additional significant contribution to their squared masses, \ie,
$m^2_{\tilde l}\approx m^2_{1/2}(c_{\tilde l}+\xi^2_0)$. This can be seen
for example by comparing \fVI{b} (solid line) with \fV{b} (dashed line).

There are two phenomenological cuts which affect the left boundary: the
chargino cut comes first and rules out small values of $\mu$ (recall that
$\mu=0$ on the left boundary), effectively shifting the left boundary to the
right by a significant amount for small $m_{1/2}$; for example, compare
\fVI{a} (solid) with \fV{a} (dashed) and \fVI{b} (solid) with \fV{b} (dashed).
Once this is satisfied, the $Z$-width cut shifts the left boundary slightly
further to the right. The top boundary (where $m^2_A=0$) is constrained a
little by the one-loop Higgs mass cut. This cut also constrains the bottom
boundary for values of $\tan\beta$ very close to 1. For $\xi_0=0$ the neutral
LSP cut is very effective on the bottom boundary, basically eliminating the
range $1\le\tan\beta\lsim2$.

The boundaries shown in \fVI{} appear in the
sequence: $(0,-1)\to(0,0)\to(0,1)$ and $(1,-1)\to(1,0)\to(1,1)$
exemplifying the motion of the left boundary to the right due to the effects
of $\xi_0$ and $\xi_A$, as described in Sec. 5.2. Other values of $\xi_0,\xi_A$
yield boundaries whose shape can be inferred from the given ones.

The one-loop SISM boundaries differ from the SSM ones in two respects: (i) the
LSP cut is less effective when applicable, and (ii) the left boundaries
are shifted to the right relative to those for the SSM. In \fVII{} we
show two illustrative cases plus one-loop contours of $\mu$ (for $\mu>0$).
(The $\Delta\mu$ shifts here follow the same pattern as in the SSM case.)
Both of these effects can be traced back to a lower value of $\mu$ for
corresponding points in parameter space (\eg, compare the $\mu$ contours in
\fVII{} with those in \fV{}).
Indeed, if $\mu$ is lower then the $\tilde\tau_1$ mass
will not be shifted downwards as much and the LSP cut will be less effective.
Also, $\mu^2$ is driven to negative values sooner and therefore one needs
to increase $m_t$ to compensate, \ie, the left boundary shifts to the right.

\newsec{Predictions for particle masses}
The spectrum of sparticle and Higgs boson masses in the SSM/SISM depends on
the particular point in the five-dimensional parameter space, and therefore it
is a complicated matter to give
the values of these masses for all allowed points in this space. However, not
all masses depend on all parameters and some of their dependences can be
worked out analytically. We now discuss the predictions for the squark,
slepton,
neutralino, chargino, and one-loop corrected Higgs boson masses. When relevant,
we also comment on the shifts on the $\mu$- and/or $B$-dependent masses which
result from the minimization of the one-loop versus tree-level Higgs
potentials.
\subsec{Squarks and sleptons}
The first and second generation (and third generation for the sneutrino) squark
and slepton masses can be determined analytically as follows (see \eg, \LN)
\eqn\Mi{m^2_i=m^2_{1/2}(c_i+\xi^2_0)-d_i{\tan^2\beta-1\over\tan^2\beta+1}M^2_W,}
with $d_i=(T_{3i}-Q)\tan^2\theta_w+T_{3i}$ (\eg, $d_{\tilde e_L}=
-{1\over2}+{1\over2}\tan^2\theta_w$ and $d_{\tilde e_R}=-\tan^2\theta_w$)
and the $c_i$ are coefficients directly calculable in terms of the gauge
couplings (see \eg, \KLNQ).
These are listed in Table III for both SSM and SISM. In detailed
calculations involving the sparticle masses, the $c$'s should be renormalized
at the physical sparticle mass \spart. However, for the purposes of this
paper, we have renormalized all the $c$'s at $M_Z$ for simplicity. We should
point out that the squark (slepton) coefficients have a $\approx10\%$
($\lsim2\%$) uncertainty due to the present uncertainties on the low-energy
gauge couplings \SISM. The gluino mass is analogously given by $m_{\tilde g}=
c_{\tilde g}m_{1/2}$, with $\approx8\%$ uncertainty on $c_{\tilde g}$. In
\fVIII{} we have plotted the ratio of these masses to the gluino mass as a
function of the gluino mass for several values of $\xi_0$ in the SSM. These
masses depend on the other parameters of the model and/or tree-level versus
one-loop minimization only to the extent that specific values of these
parameters may exclude certain values of $\tan\beta,\xi_0$, and $m_{1/2}$. In
fact, in \fVIII{} (where we have taken $\xi_A=0$, $\mu>0$, and $m_t=125\GeV$)
not all lines start at $m_{\tilde g}\approx150\GeV$ or continue up to
$m_{\tilde g}=1\TeV$, due to the various cuts on the paramater space. This
dependence is particularly important for the $\xi_0=0$ case, where the upper
bound on the masses varies quite a bit with $m_t$. Note that the near
proportionality to $m_{\tilde g}$ is only broken by small D-term
($\tan\beta$-dependent) effects. In fact, to a good approximation (which
improves with increasing $m_{\tilde g}$) these ratios are simply given by
\eqn\Mii{{m_i\over m_{\tilde g}}\approx{\sqrt{c_i+\xi^2_0}\over c_{\tilde g}},}
with the $c_i$ given in Table III. Both squarks and sleptons can be lighter
or heavier than the gluino, depending on the value of $\xi_0$. However,
squarks are always heavier than sleptons.

All the above remarks apply to the SISM as well. With the help of Eq. \Mii\
and Table III one can easily reproduce the analog of \fVIII{} for the SISM.
Note that the gluino and all of the squarks and sleptons are lighter in the
SISM than in the SSM for the same point in parameter space \SISM, although
the ratios $m_i/m_{\tilde g}$ are larger in the SISM than in the SSM.

The $\tilde\tau_{1,2},\tilde b_{1,2},\tilde t_{1,2}$ mass eigenstates receive
additional contributions from the off-diagonal left-right mixing term (plus
the corresponding fermion masses). In the case of $\tilde b_{1,2}$ we have
verified that these eigenstates have masses very approximately equal to that
of $\tilde d_R$ and $\tilde d_L$ respectively. This also happens for the
stau mass eigenstates which are close to $\tilde e_{R,L}$, although the
discrepancy grows with $\xi_0$ since then larger values of $\mu$ can occur.
A rather interesting effect occurs in this case which is worth pointing out.
{}From Eq. \Mi, $d_{\tilde e_R}<0$ and therefore $m_{\tilde e_R}$ grows with
increasing $\tan\beta$. In \fIX{} (for $\mu>0$) we show contours of
$m_{\tilde\tau_1}$, which exhibit just the opposite $\tan\beta$ dependence,
due to the off-diagonal $m_\tau(A_\tau+\mu\tan\beta)$ term. Thus, the effect
of the D-term is completely washed out by the $\lambda_\tau$ contribution.

The $\tilde t_1$ mass eigenstate can be significantly lighter than the average
squark mass. In \fIX{} we indicate the ranges of $m_{\tilde t_1}$ throughout
each of the boundaries shown. For example, for $m_{\tilde g}=415\,(690)\GeV$
(\ie, $m_{1/2}=150\,(250)\GeV$) and $\xi_0=\xi_A=0$, the $\tilde u_R$ squark
mass is $\approx360\,(600)\GeV$ whereas $\tilde t_1:270-310\,(495-545)\GeV$
for $\mu>0$ and $240-285\,(425-520)\GeV$ for $\mu<0$. This discrepancy grows
with $m_{1/2}$ since $\mu$ scales with $m_{1/2}$. The lowest value of
$m_{\tilde t_1}$ occurs for $\xi_0\approx1$ since then smaller values of
$m_{1/2}$ are allowed. For $m_{\tilde g}=150\GeV$ and $\xi_A=-1,-2,-3$, we
obtain $m_{\tilde t_1}\gsim100,80,45\GeV$. However, this effect occurs only
for $m_t\approx110\GeV$ and $\tan\beta\approx2$, \ie, a rather fine-tuned and
very small region of parameter space. A more typical lower bound would be
$m_{\tilde t_1}\gsim100\GeV$.
\subsec{Neutralinos and charginos}
The neutralino $\tilde\chi^0_i\,(i=1,2,3,4)$ and chargino $\tilde\chi^\pm_i\,
(i=1,2)$ masses depend on just three parameters: $m_{1/2},\mu,\tan\beta$.
We have explored the masses of the lightest of each kind (using $\mu_{loop}$)
and found that they get shifted relative to what would be obtained using
$\mu_{tree}$. As discussed in Sec. 5.2, the $\Delta\mu$ shift is largest for
small $\mu$. This reflects itself on the shifts of the $\tilde\chi^0_1$ and
$\tilde\chi^\pm_1$ masses as well. For example, for the $\xi_0=\xi_A=0$ case
for $m_{1/2}=150\GeV$ we get $m^{tree}_{\wt\chi^0_1}=50\,(55)\GeV\to
m^{loop}_{\wt\chi^0_1}=41-66\,(50-53)\GeV$ and
$m^{tree}_{\wt\chi^\pm_1}=75\,(100)
\GeV\to m^{loop}_{\wt\chi^\pm_1}=58-64\,(92-93)\GeV$. Since $\Delta\mu<0$, we
expect the shifts to be negative, as observed. Note that the mass shifts are
rather small ($\lsim$few GeV) unless one is very close to the left boundary.
Furthermore, as one moves away from this boundary and $\mu$ increases, its
effect on the mases decreases considerably, making the $\Delta\mu$ shift
irrelevant.

\subsec{Higgs bosons}
In both the SSM and SISM models we consider here, the Higgs sector is
comprised of two complex Higgs doublets. After electroweak symmetry breaking
there remain three neutral (two scalars $h,H$ and a pseudoscalar $A$) and one
charged $(H^\pm)$ physical Higgs bosons. Depending on the specific point in the
allowed parameter space, the mass of $h$ (and to a lesser extent $H,A,H^\pm$)
may receive significant one-loop radiative corrections
\refs{\okada,\ERZI,\DNh,\Higgs}. Here we extend
our previous `no-scale' analysis ($\xi_0=\xi_A=0$) \Higgs\ of the one-loop
corrected Higgs boson masses in the allowed $(m_t,\tan\beta,m_0,m_{1/2},A)$
parameter space by considering other $\xi_0,\xi_A$ options as well. As a
fine-tuning constraint (see Sec. 3.3), we restrict $m_{1/2}<400\GeV$ which
effectively bounds the various Higgs masses from above. The analysis of the
MSSM Higgs sector is usually parameterized by $\tan\beta$ and $m_A$;
in contrast, for the SSM/SISM unified models $m_A$ is redundant,
since $m_A$ (as well as the other Higgs masses) are {\it predicted}
from the original set of five parameters.

In order to include radiative corrections to the Higgs boson masses,
simplifying choices for the many low-energy parameters are usually made (none
of which are made in this paper). For
example, it is typical to assume $A_t=A_b=\mu$ and
$m^2_Q=m^2_U=m^2_D=m^2_{\tilde q}$. However, as discussed in the Introduction,
the assumption of universal soft supersymmetry breaking at $M_U$ in the SSM
and SISM leads to a correlation of the many low-energy parameters, and these
correlations are essential to the overall predictiveness of the models. The
Higgs spectrum we present here incorporates these crucial low-energy
correlations.

As expected, the Higgs masses scale with $m_{1/2}$, and the one-loop radiative
corrections depend strongly on the value of $m_t$. In \fVII{} (SISM) and \fIX{}
(SSM) we give contours of $h$ ($\mu<0$ boundaries; contours for $\mu>0$ are
very
similar). One can see that $h$ is driven quickly to values above $M_Z$,
although
small corners of parameter space exist where it can be as light as
$\sim {1\over 2}M_Z$. For the `no-scale' scenario (\fVII{a} and \fIX{a}), the
maximal values for $h$ are $m_h^{max}\simeq 120\,(130)\GeV$ for the SSM (SISM).
The larger $m_h^{max}$ value for SISM is due to the less restrictive LSP
cut as discussed in Sec. 5.3. For a {\it given} point in the parameter space
we find that overall there is a slight systematic downward shift ($<5\%$) in
$m_h$ in the SISM compared to the SSM. This is probably a consequence of the
slightly smaller values of $\mu$ in the SISM relative to the SSM (see Sec.
5.3).
For increasing values of $\xi_0$, the
values of $m_h^{max}$ rise and saturate with the fine-tuning or ``naturalness"
value of $m_{1/2}=400\GeV$; the value of $m_h^{max}\simeq 135\,(130)\GeV$ for
SSM (SISM) and for both $(\xi_0,\xi_A)=(1,1),(1,-1)$ choices. The values of
$m_h^{max}$ for the $\mu<0$ cases are similar.

{}From the fine-tuning constraints and the various constraints on the
$(m_t,\tan\beta)$ parameter space discussed at length in Sec. 5.2, we can make
the prediction that $m_h<135\,(130)\GeV$ in `natural' SSM (SISM) scenarios
irrespective of the supersymmetry breaking choice; this is a rather restrictive
and striking prediction. This result agrees with the previous analysis of
Drees and Nojiri \DNh, even though they made the additional assumption of
$A=B+m_0$. In comparison, the MSSM analysis (see fig. 1a of Ref. \KZ) shows
that $m_h^{max}\simeq 132$ (for $m_{\wt q}=1\TeV$), again in close agreement
with our result.

Due to the more severe LSP phenomenological cuts,
the `no-scale' SSM seems to be rather unique in requiring a `light' $h$
($m_h<120\GeV$). Of course, if the fine-tuning constraint were
relaxed and $m_{1/2}$ increased, then $m_h^{max}$ would rise as well.
These mass constraints suggest that if $h$ is not found at LEP,
the $WhX\rightarrow l\gamma\gamma X$ final states at SSC/LHC may be the
only hope of discovering the $h$ Higgs.
For a more detailed phenomenological discussion of the $h$ Higgs
boson, including radiative corrections, see Refs. \refs{\KZ,\Barger}.

As for the $h$ Higgs, as $m_{1/2}$ increases, the masses for $H,A,H^\pm$ rise
as well, and become increasingly degenerate. For $m_{1/2}=400\GeV$, in \fX{} we
present contours of $m_A$ for the $(\xi_0,\xi_A)=(1,\pm1)$ cases for both SSM
and SISM, and $\mu>0$. Due to the high degree of degeneracy, the contours can
be assumed to also represent contours of $m_H$ and $m_{H^\pm}$ to within $5\%$.

In summary, the Higgs sector of `natural' supergravity models generically
predict a relatively light Higgs scalar, $m_h<135\GeV$ possibly detectable at
the SSC/LHC. This limit takes into account naturalness, the full set of
constraints on the allowed parameter space, and is maintained over a wide range
of supersymmetry breaking scenarios. In this limit, the $H,A,H^\pm$ Higgs
masses
are nearly degenerate, but could be anywhere from $100\GeV-1\TeV$. Due to the
near degeneracy of the SSM/SISM Higgs sectors, it is unlikely that discovery of
a minimal supersymmetry Higgs sector will allow for an immediate discrimination
between these two models.

\subsec{Sample particle mass ranges}
In Tables IV and V we have collected the mass ranges for all the particle
species in the SSM (results for the SISM are similar) when one sweeps the
entire one-loop allowed region in $(m_t,\tan\beta)$
for the given $\xi_0,\xi_A$, and $m_{1/2}$ values. The values of $m_{1/2}$ have
been chosen such that the corresponding regions (in $(m_t,\tan\beta)$) are
not too restrictive. For example, for too small values of $m_{1/2}$ the regions
tend to be confined to small values of $\tan\beta$. Thus, we avoid `small'
regions of parameter space in these two variables. For the $\xi_0=\xi_A=0$ case
(in Table IV), $m_{1/2}=110\GeV$ is close to the lowest possibly obtainable
value ($\approx100\GeV$). The mass ranges for this case should be compared
with their tree-level counterparts given in Ref. \KLNPY. The ranges have some
differences but they are quite close, except for the Higgs masses which were
not given in Ref. \KLNPY.
For the $\xi_0=1,\xi_A=0$ case (in Table V) the lowest
value of $m_{1/2}$ for which `large' regions still exist depends on the sign of
$\mu$, as indicated in the Table. The most striking difference between this
case (and in general for any value of $\xi_0\not=0$) is that values of the
squark and gluino masses as low as their current experimental lower bounds are
attainable (although not shown explicitly on the table since these correspond
to `small' regions of parameter space).
Note also that a light chargino ($\wt\chi^\pm_1$) is always allowed, although
the fraction of the parameter space for which this happens decreases with
increasing $m_{1/2}$.

\newsec{Conclusions}
The study of supersymmetric models is a subject of great topical interest
since this is the physics beyond the standard model which is most likely to
exist on theoretical grounds. The minimal supersymmetric extension of the
standard model (MSSM) by itself contains so many parameters (21) that
comprehensive studies are impractical. In fact, to alleviate this problem
people routinely make simplifying assumptions about the various parameters,
typically assuming a sort of `low-energy unification' of the sparticle
masses.

The approach we advocate does away with ad-hoc low-energy ans\"atze by
invoking the structure of unified models where well motivated theoretical
constraints reduce the dimension of the parameter space down to eight. However,
even these constraints are not enough to produce a consistent picture of
low-energy physics since the electroweak symmetry must be broken as well. This
last constraint determines $\mu$ and $B$ and generally correlates them with the
dominant source of supersymmetry breaking. The major advantage of the above
constraints is to reduce the dimension of the parameter space down to just
five.

Our present study utilizes the one-loop effective Higgs potential to execute
the last step of radiative electroweak symmetry breaking. This makes the
results largely independent of the renormalization scale $Q$ used (\ie, the
scale where the RGEs are stopped and the Higgs potential is minimized). The
use of this potential is a technical advancement which has yet to catch up
in the literature, and we believe it to be essential in the determination of
the ground state of this class of supersymmetric unified models.
We have explored the virtues of this more sophisticated
approach and in the process have realized that a Higgs-field--independent
term in $\Delta V$ ruins the $Q$-independence of $V_1$. For fixed-$Q$
calculations this term is irrelevant since it drops out in the minimization
process. However, the whole point of using $V_1$ is precisely to obtain
$Q$-independence. We have shown explicitly that if this piece is subtracted
out, then $V_1$ is indeed $Q$-independent to one-loop order.

We have also discussed the `fine-tuning constraint' in the light of our
analysis of the parameter space. We found that a good measure of this effect
is given by coefficients $c\propto m^2_{1/2}(a+b\xi^2_0)$ which can easily
surpass a `limit' $\Delta=50-100$ if $m_{1/2}>400\GeV$ for all values of
$\xi_0$. Our results here are not new, we still get $m_{\tilde g,\tilde q}\lsim
1\TeV$. However, these coefficients help quantify the level of `un-naturalness'
in a model. For example, we have found that in the minimal $SU(5)$ supergravity
GUT, proton decay constraints can be satisified only for values of the
parameters which give $c\gsim100$. Thus indicating a possible fine-tuning
problem in this model.

Our study has been based on two $SU(3)\times SU(2)\times U(1)$ supersymmetric
models which exemplify the low-energy limits of traditional (SSM) and
string-inspired (SISM) unified models. For these models we have described in
detail the completely bounded regions in ($m_t,\tan\beta$) space and their
dependence on the other parameters of the model. We have also compared the
tree-level and one-loop values of $\mu$ and concluded that
$|\Delta\mu|=|\mu_{loop}-\mu_{tree}|$ is largest for small $\mu$, as expected.
In fact, the relative shift in $\mu$ is maximal ($-100\%$) at the one-loop
left boundary and decreases as $m_t$ increases. Fortunately, this small
region of parameter space, where one-loop perturbation theory is unreliable,
is excluded on phenomenological grounds.

We have studied the sparticle spectrum and have given useful plots
of the first and second generation squark and slepton masses. We have shown
that the inclusion of $\lambda_\tau\not=0$ is relevant since $\mu$ effects
dominate over the D-term contributions to the $\tilde\tau_{1,2}$ masses.
The lightest stop eigenstate $\tilde t_1$ is generally lighter than the
average squark mass, and can even be as light as ${1\over2}M_Z$ although
only for a very small region of parameter space; a more typical lower bound
is $m_{\tilde t_1}\gsim100\GeV$. We have also studied the one-loop corrected
Higgs boson masses and concluded that for `natural' values of the parameters
(\ie, $m_{\tilde g,\tilde q}<1\TeV$) one must have $m_h<135\GeV$. Finally,
we have given mass ranges for all particle species for typical values of
the parameters.
If supergravity is indeed realized in Nature, then the correlations among the
many sparticle and Higgs boson masses that we have presented here should be
dramatically revealed in the near future.

\bigskip
\bigskip
\bigskip
\noindent{\it Acknowledgments}: This work has been supported in part by DOE
grant DE-FG05-91-ER-40633. The work of J.L. has been supported in part by an
ICSC-World Laboratory Scholarship.The work of D.V.N. has been supported in part
by a grant from Conoco Inc. The work of K.Y. was supported in part by the Texas
National Laboratory Research Commission under Grant No. RCFY9155, and in part
by the U.S. Department of Energy under Grant No. DE-FG05-84ER40141. We would
like to thank Doni Branch and the HARC Supercomputer Center for the use of
their
NEC SX-3 supercomputer.
\vfill\eject

\appendix{A}{Calculation of $\Delta V$}
In this appendix we present some details of the calculation of $\Delta V$ used
throughout our work. The evaluation of $\Delta V$ proceeds from the following
definition
\eqn\Ai{\Delta V={1\over64\pi^2}{\rm Str}\,{\cal M}^4
\left(\ln{{\cal M}^2\over Q^2}-{3\over2}\right),}
where ${\rm STr}\,f({\cal M}^2)=\sum_{i}C_i(-1)^{2j_i}(2j_i+1)f(m^2_i)$
and $m^2_i$ are the Higgs-field--dependent mass-squared eigenstates with spin
$j_i$. We refer the reader to the literature for the necessary squark
\refs{\LN,\BERZ}, slepton \refs{\LN,\BERZ}, Higgs \GH, and gaugino/Higgsino
\LNY\ mass formulas in the MSSM. The factor $C_i$ that enters into the
supertrace accounts for the color degrees of freedom, and we define a spin
factor $S_i=(-1)^{2j_i}(2j_i+1)$, where $j_i$ is the spin of the $i$th
particle. The appropriate $S_i,C_i$ factors for the particles which contribute
nonnegligibly to $\Delta V$ are summarized in Table VI.

In calculating charged Higgs masses, scalar field values away from the minimum
(and which therefore break $U_{em}(1)$ invariance) are needed. In this case
charge-conjugate particles are no longer degenerate in mass and we treat each
charge-conjugate state as a separate mass-eigenstate in the supertrace.
For example, for the squarks $\wt q$, the complete set of mass eigenstates
(at the minimum of the potential) would be
\eqn\Aii{\wt q=({\wt u_{iL}},{\wt u^*_{iL}},{\wt u_{iR}},{\wt u^*_{iR}},
{\wt d_{iL}},{\wt d^*_{iL}},{\wt d_{iR}},{\wt d^*_{iR}},
i=1,2,\quad {\wt t_{1}},{\wt t^*_{1}},{\wt t_{2}},{\wt t^*_{2}},{\wt b_{1}},
{\wt b^*_{1}},{\wt b_{2}},{\wt b^*_{2}}),}
where $i=1,2$ labels the first and second generation squarks.

The contributions of the $u,d,c,s,e,\mu,\tau$ fermions to $\Delta V$ are
utterly negligible (and would vanish identically in the massless limit),
and we do not include them in our calculations. Depending on whether the
scalar field configuration is chosen to be at (i) the origin of field space
or (ii) away from it, the contributions to $\Delta V$ from the two Higgs
doublets and the $W^\pm$ and $Z$ fields are included as follows:
(i) the $W^\pm$ and $Z$ fields are massless and all eight Higgs mass
eigenstates must be included (\ie, the Goldstone bosons are not massless);
(ii) the $W^\pm$ and $Z$ fields are massive and only the Higgs physical states
$h,A,H,H^\pm$ are included (\ie, the Goldstone bosons are massless). In
certain cases, it is possible for $m^2_i<0$, and in this event we make the
replacement $m^2_i\rightarrow |m^2_i|$ \GRZ.

We thus calculate $\Delta V$ using the following complete supertrace formula
\eqna\Aiii
$$\eqalignno{{\rm STr}f({\cal M}^2)=&3\sum_{i=1,24}^{\wt q}
f(m^2_{\wt q_i})-6\sum_{i=1,4}^{q}f(m^2_{q_i})
+\sum_{i=1,18}^{\wt l}f(m^2_{\wt l_i})\cr
&+\sum_{i=1,8}^{H}f(m^2_{H_i})
-2\sum_{i=1,8}^{\wt \chi}f(m^2_{\chi_i})+3\sum_{i=1,3}^{W,Z}f(m^2_{V_i})\cr
&-16f(m^2_{\wt g}).&\Aiii{}\cr}$$

\listrefs
\input tables
\centerjust
\vbox{\tenpoint\noindent {\bf Table I}: Values for the fine-tuning coefficients
corresponding to typical points in parameter space (with $\tan\beta=1.73$,
$m_t=125\GeV$, $\mu>0$, $A_t=-0.6m_0$ \AN) which satisfy proton decay bounds
in minimal $SU(5)$ supergravity models for $M_H=3M_U$. The $m_{1/2}$ values are
in GeV.}
\medskip
\thicksize=1.0pt
\begintable
$m_{1/2}$|$\xi^{min}_0$|$\xi_A$|$c_\mu$|$c_t$\cr
74|8.1|$-5.4$ |39|113 \nr
122|6.5|$-3.2$ |67|204 \nr
187|5.4|$-1.6$ |107 |342 \nr
267|4.5|$-.41$ |155 |503 \nr
364|3.8|$+.55$ |217 |699 \endtable
\bigskip
\bigskip
\vbox{\tenpoint\noindent {\bf Table II}: Tree-level and one-loop values
for $\mu$ contours in \fV{a} ($\xi_0=\xi_A=0$) for $\mu>0$ (similar results
for $\mu<0$). All masses in GeV.}
\medskip
\begintable
\multispan{3} \tstrut
$m_{1/2}=150\GeV$ \|\multispan{3} $m_{1/2}=250\GeV$ \cr
$\mu_{tree}$|$\mu_{loop}$|$\Delta\mu$\|$\mu_{tree}$|$\mu_{loop}$|$\Delta\mu$\cr
$75$|$25$|$-50$\|$150$|$60$|$-90$\nr
$100$|$65$|$-35$\|$200$|$130$|$-70$\nr
$150$|$120$|$-30$\|$300$|$235$|$-65$\nr
$200$|$170$|$-30$\|$350$|$285$|$-65$\nr
$240$|$210$|$-30$\|$400$|$335$|$-65$\endtable
\bigskip
\bigskip
\vbox{\tenpoint\noindent {\bf Table III}: Values of the $c_i$ coefficients
in Eq. \Mi\ for the first and second generation sfermions (\ie,
$c_{\tilde e_R}=c_{\tilde\mu_R}$, and so on) for the SSM and SISM. Also
shown is $c_{\tilde g}\equiv m_{\tilde g}/m_{1/2}$.}
\smallskip
\begintable
$c_i$|SISM |SSM \cr
$c_{\tilde u_L,\tilde d_L}$|3.91 |6.28 \nr
$c_{\tilde u_R}$|3.60 |5.87 \nr
$c_{\tilde d_R}$|3.55 |5.82 \nr
$c_{\tilde e_L,\tilde\nu}$|0.402|0.512\nr
$c_{\tilde e_R}$|0.143|0.149\nr
$c_{\tilde g}$|2.01 |2.77 \endtable
\vfill\eject
\vbox{\tenpoint\noindent {\bf Table IV}: Particle mass ranges in the SSM
(throughout the allowed $(m_t,\tan\beta)$ region) for $\xi_0=\xi_A=0$ and a
sample of $m_{1/2}$ values. The $\pm$ indicate the sign of $\mu$. The lowest
value of $m_{1/2}$ is close its absolute minimum. All masses in GeV.}
\bigskip
\begintable
|\multispan{6}\hfill$\xi_0=0,\xi_A=0$\hfill\cr
$m_{1/2}$|$110^+$|$110^-$|$150^+$|$150^-$|$250^+$|$250^-$\cr
$\wt\chi^0_1$ |$ 34- 50$|$ 22- 38$|$ 36- 67$|$ 32- 58$|$ 37-104$|$ 60-102$\nr
$\wt\chi^0_2$ |$ 55-103$|$ 62- 69$|$ 73-133$|$ 78-104$|$ 76-175$|$112-188$\nr
$\wt\chi^0_3$ |$ 93-204$|$123-196$|$ 87-274$|$104-269$|$116-234$|$122-440$\nr
$\wt\chi^0_4$ |$136-220$|$168-240$|$161-287$|$179-305$|$231-269$|$242-466$\nr
$\wt\chi^\pm_1$ |$ 53-103$|$ 45- 65$|$ 56-133$|$ 52-102$|$ 53-176$|$ 81-187$\nr
$\wt\chi^\pm_2$ |$137-220$|$170-235$|$161-287$|$180-301$|$230-272$|$242-462$\nr
$\tilde\nu$   |$ 45- 60$|$ 45- 60$|$ 86- 94$|$ 86- 94$|$167-171$|$167-171$\nr
$\tilde e_L$  |$ 87- 92$|$ 87- 92$|$114-117$|$114-117$|$183-185$|$183-185$\nr
$\tilde e_R$  |$ 55- 61$|$ 55- 61$|$ 68- 73$|$ 68- 73$|$103-106$|$103-106$\nr
$\tilde\tau_1$|$ 43- 59$|$ 43- 57$|$ 53- 72$|$ 49- 71$|$ 91-105$|$ 89-104$\nr
$\tilde\tau_2$|$ 87-100$|$ 88-101$|$114-124$|$114-126$|$183-187$|$183-188$\nr
$\tilde g$|$300$|$300$|$415$|$415$|$690$|$690$\nr
$\tilde u_L$  |$271-272$|$271-272$|$372-374$|$372-374$|$624-625$|$624-625$\nr
$\tilde u_R$  |$264-265$|$264-265$|$362-362$|$362-362$|$605-605$|$605-605$\nr
$\tilde d_L$  |$280-282$|$280-282$|$379-381$|$379-381$|$628-629$|$628-629$\nr
$\tilde d_R$  |$266-267$|$266-267$|$362-363$|$362-363$|$604-604$|$604-604$\nr
$\tilde b_1$  |$247-266$|$242-258$|$336-362$|$328-358$|$565-603$|$558-597$\nr
$\tilde b_2$  |$266-273$|$268-275$|$361-371$|$364-370$|$592-614$|$594-608$\nr
$\tilde t_1$  |$189-252$|$165-220$|$269-316$|$242-284$|$494-545$|$425-521$\nr
$\tilde t_2$  |$313-346$|$338-360$|$391-425$|$417-440$|$643-651$|$638-656$\nr
$h$           |$ 42-102$|$ 66-104$|$ 46-109$|$ 46-112$|$ 46- 97$|$ 46-121$\nr
$A$           |$ 48-232$|$ 86-229$|$ 49-325$|$ 49-321$|$ 53-278$|$ 54-539$\nr
$H$           |$ 96-244$|$ 99-240$|$ 93-332$|$ 94-329$|$ 93-277$|$ 94-542$\nr
$H^\pm$         |$ 94-244$|$118-241$|$ 94-333$|$ 95-330$|$ 97-289$|$ 98-545$
                                                                \endtable
\vfill\eject
\vbox{\tenpoint\noindent {\bf Table V}: Particle mass ranges in the SSM
(throughout the allowed $(m_t,\tan\beta)$ region) for $\xi_0=1,\xi_A=0$ and a
sample of $m_{1/2}$ values. The $\pm$ indicate the sign of $\mu$. The lowest
value of $m_{1/2}$ (for the given sign of $\mu$) is close its absolute minimum.
All masses in GeV.}
\bigskip
\begintable
|\multispan{6}\hfill$\xi_0=1,\xi_A=0$\hfill\cr
$m_{1/2}$|$70^+$|$85^-$|$150^+$|$150^-$|$250^+$|$250^-$\cr
$\wt\chi^0_1$ |$ 27- 34$|$ 26- 29$|$ 36- 67$|$ 23- 60$|$ 38-108$|$ 31-104$\nr
$\wt\chi^0_2$ |$ 40- 75$|$ 50- 52$|$ 73-133$|$ 78-108$|$ 78-212$|$ 82-201$\nr
$\wt\chi^0_3$ |$101-147$|$136-150$|$ 84-877$|$104-304$|$115-568$|$119-332$\nr
$\wt\chi^0_4$ |$113-168$|$165-179$|$162-885$|$178-336$|$232-573$|$238-340$\nr
$\wt\chi^\pm_1$ |$ 45- 76$|$ 45- 48$|$ 56-133$|$ 46-107$|$ 55-212$|$ 50-201$\nr
$\wt\chi^\pm_2$ |$119-168$|$169-182$|$161-883$|$179-331$|$232-571$|$238-338$\nr
$\tilde\nu$   |$ 58- 69$|$ 82- 83$|$173-183$|$173-177$|$301-307$|$301-307$\nr
$\tilde e_L$  |$ 94- 98$|$115-115$|$185-190$|$188-190$|$308-311$|$308-311$\nr
$\tilde e_R$  |$ 83- 87$|$101-101$|$161-167$|$165-167$|$268-272$|$268-272$\nr
$\tilde\tau_1$|$ 79- 85$|$ 45- 91$|$ 85-166$|$ 85-165$|$172-271$|$172-271$\nr
$\tilde\tau_2$|$ 94-103$|$121-135$|$186-208$|$188-208$|$303-318$|$303-318$\nr
$\tilde g$|$195$|$235$|$415$|$415$|$690$|$690$\nr
$\tilde u_L$  |$181-184$|$223-223$|$401-404$|$401-403$|$673-675$|$673-675$\nr
$\tilde u_R$  |$180-181$|$220-220$|$392-393$|$392-392$|$654-655$|$654-655$\nr
$\tilde d_L$  |$195-198$|$237-237$|$405-409$|$408-409$|$675-677$|$675-677$\nr
$\tilde d_R$  |$184-185$|$223-223$|$392-393$|$392-393$|$653-653$|$653-653$\nr
$\tilde b_1$  |$168-183$|$143-197$|$264-390$|$260-385$|$464-648$|$460-647$\nr
$\tilde b_2$  |$184-188$|$230-234$|$371-394$|$374-394$|$578-654$|$581-654$\nr
$\tilde t_1$  |$140-204$|$182-197$|$204-324$|$245-303$|$385-567$|$310-559$\nr
$\tilde t_2$  |$243-279$|$300-312$|$401-453$|$416-452$|$608-688$|$610-743$\nr
$h$           |$ 47- 92$|$ 46-100$|$ 41-112$|$ 45-113$|$ 38-124$|$ 39-125$\nr
$A$           |$ 88-170$|$ 48-152$|$ 50-244$|$ 50-397$|$ 55-240$|$ 55-897$\nr
$H$           |$108-187$|$ 99-152$|$ 97-234$|$ 98-402$|$ 99-214$|$ 99-883$\nr
$H^\pm$         |$118-187$|$ 97-171$|$ 97-246$|$ 98-404$|$100-241$|$100-899$
                                                                \endtable
\vfill\eject
\vbox{\tenpoint\noindent {\bf Table VI}: Values for the multiplicity, spin,
and $C_i$ and $S_i$ factors for each of the particles of each species
that have been included in the supertrace for $\Delta V$. The index $i$ runs
from 1 to $n$. We have not included the $u,d,c,s,e,\mu,\tau$ fermions since
their contribution is negligible.}
\bigskip
\thicksize=1.0pt
\begintable
particle |$n$ |Spin ($j$) |$C_i$ |$S_i$ \cr
$\wt q_i$|$24$ |$0$|$3$|$1$\nr
$\wt l_i$|$18$ |$0$|$1$|$1$\nr
$H_i$|$8$|$0$|$1$|$1$\nr
$q_i$|$4$|${1\over2}$|$3$|$-2$ \nr
$\wt \chi_i$|$8$|${1\over2}$|$1$|$-2$ \nr
$\wt g$|$1$|${1\over2}$|$8$|$-2$ \nr
$W$|$2$|$1$|$1$|$3$\nr
$Z$|$1$|$1$|$1$|$3$\endtable

\listfigs
\bye